\documentclass[journal=apchd5,manuscript=article]{achemso}

\usepackage[version=3]{mhchem} 
\usepackage{siunitx}[=v2]
\usepackage{xcolor}
\mciteErrorOnUnknownfalse
\usepackage[utf8]{inputenc} 
\DeclareUnicodeCharacter{2009}{\,} 

\pdfoutput=1

\author{Chiara Lombardo}
\affiliation[ETH Zurich]{Nanophotonic Systems Laboratory, Department of Mechanical and Process Engineering, ETH Zurich, Tannenstrasse 3, 8092 Zurich, Switzerland}
\author{Andrea Sottini}
\affiliation[ETH Zurich]
{Nanophotonic Systems Laboratory, Department of Mechanical and Process Engineering, ETH Zurich, Tannenstrasse 3, 8092 Zurich, Switzerland}
\author{Sarina Seiter}
\affiliation[ETH Zurich]
{Nanophotonic Systems Laboratory, Department of Mechanical and Process Engineering, ETH Zurich, Tannenstrasse 3, 8092 Zurich, Switzerland}
\author{Gerard Colas des Francs}
\affiliation{Université Bourgogne Europe, CNRS, Laboratoire Interdisciplinaire Carnot de Bourgogne ICB UMR 6303, F-21000 Dijon, France}
\author{Jaime Ortega Arroyo}
\affiliation[ETH Zurich]
{Nanophotonic Systems Laboratory, Department of Mechanical and Process Engineering, ETH Zurich, Tannenstrasse 3, 8092 Zurich, Switzerland}
\email{jarroyo@ethz.ch}
\author{Romain Quidant}
\affiliation[ETH Zurich]
{Nanophotonic Systems Laboratory, Department of Mechanical and Process Engineering, ETH Zurich, Tannenstrasse 3, 8092 Zurich, Switzerland}

\title[An \textsf{achemso} demo]
  {Leveraging partial coherence in interferometric microscopy to enhance nanoparticle detection sensitivity and throughput}
  
\abbreviations{NA, iSCAT, BFP, FOV, LED, RGG, SNR, NP}

\begin{document}
\begin{abstract}
Interferometric-based microscopies stand as powerful label-free approaches for monitoring and characterising chemical reactions and heterogeneous nanoparticle systems in real time with single particle sensitivity. Nevertheless, coherent artifacts, such as speckle and parasitic interferences, together with limited photon fluxes from spatially incoherent sources, pose an ongoing challenge in achieving both high sensitivity and throughput. In this study, we systematically characterise how partial coherence affects both the signal contrast and the background noise level; thus, it offers a route to improve the signal-to-noise ratio from single nanoparticles (NPs), irrespective of their size and composition; or the light source used. We first validate that lasers can be modified into partially coherent sources with performance matching that of spatially incoherent ones; while providing higher photon fluxes. Secondly, we demonstrate that tuning the degree of partial coherence not only enhances the detection sensitivity of both synthetic and biological NPs, but also affects how signal contrasts vary as a function of the focus position. Finally, we apply our findings to single-protein detection, confirming that these principles extend to differential imaging modalities, which deliver the highest sensitivity. Our results address a critical milestone in the detection of weakly scattering NPs in complex matrices, with wide-ranging applications in biotechnology, nanotechnology, chemical synthesis, and biosensing; ushering a new generation of microscopes that push both the sensitivity and throughput boundaries without requiring beam scanning.

\end{abstract}

\section{Introduction}
Recent advances in biotechnology, nanotechnology, material science, and chemical synthesis have enabled the engineering of new functional nanoparticle systems with applications ranging from gene delivery \cite{Pozzi2023LookingDelivery, Tian2013NanoparticlesDelivery}, targeted therapy \cite{Yetisgin2020TherapeuticApplications}, biosensing \cite{Doria2012NobleApplications} to heterogeneous catalysis \cite{Sankar2020RoleCatalysts}. Additionally, molecular profiling biological nanoparticles found in the secretome like extracellular vesicles (EVs) hold promise as next-generation liquid biopsies \cite{Zhao2019ExtracellularApplication} and drug delivery \cite{Elsharkasy2020ExtracellularHow} carriers. Consequently, there is a growing demand for high-throughput quantitative characterisation tools offering single particle sensitivity.  By eliminating the need for sample handling and processing, label-free approaches are amongst the most suited for this task. Those based on interferometry stand out due to their inherent high sensitivity. The combination of shot-noise limited detection with high-photon flux light sources ensures that enough scattered photons from the smallest of nanoparticles reach the sensor and generate sufficient signal contrast, enabling single protein and nucleic acid detection, tracking and subsequent characterisation \cite{Dahmardeh2023Self-supervised10kDa, OrtegaArroyo2014Label-freeProtein, Piliarik2014DirectSites} as well as monitoring of complex reactions ranging from autocatalysis \cite{Ortega-Arroyo2016VisualizationSystem}, nanoparticle formation\cite{Guo2025Real-timeSelf-assembly, Guo2023Real-TimeFormation}, covalent organic framework formation \cite{Gruber2024EarlyOperando} and tracking of single particle ion dynamics\cite{Merryweather2021OperandoBatteries}.   

\begin{figure}
      \centering
      \includegraphics{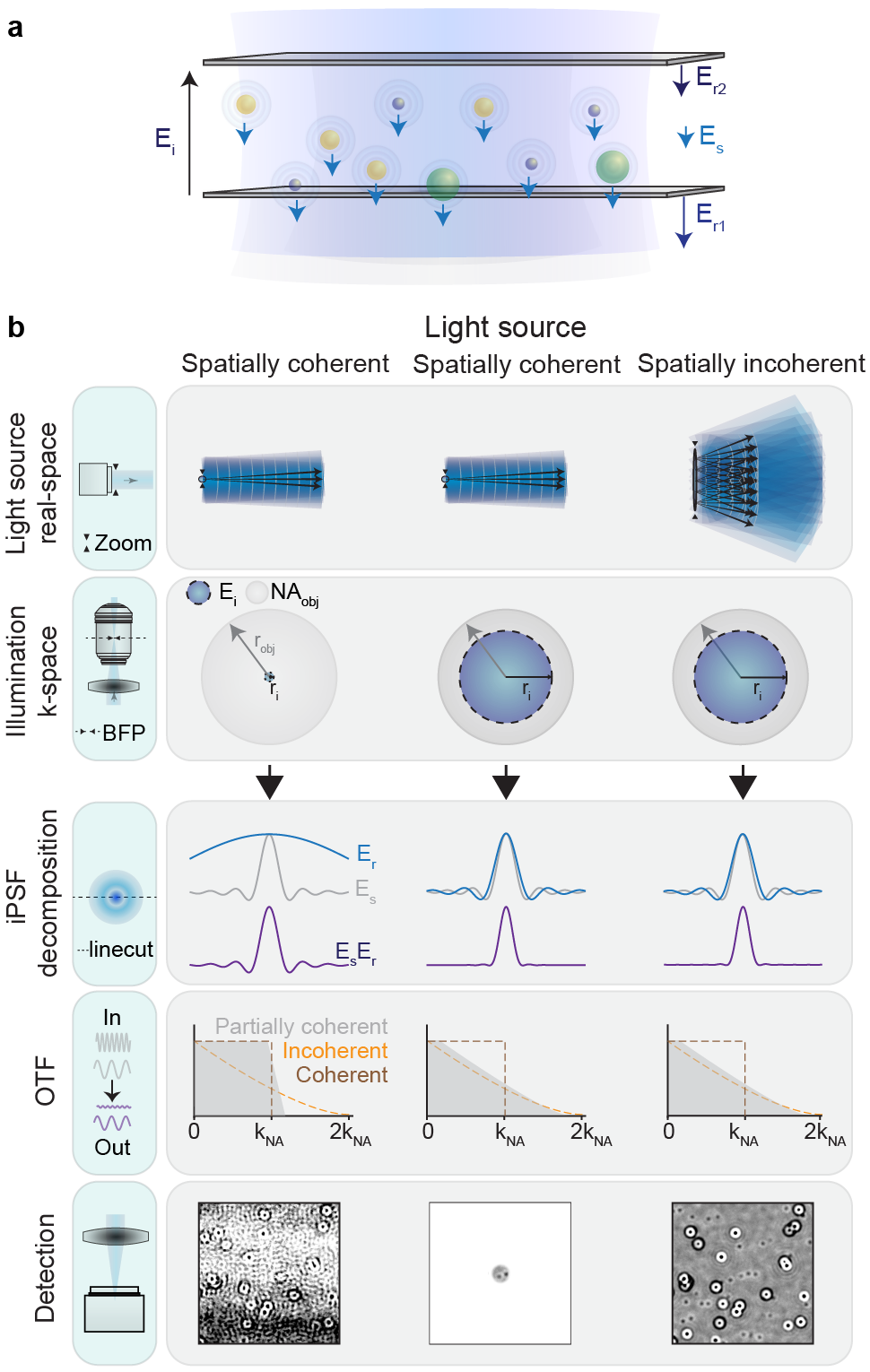}
      \caption{(a) Cartoon depicting the principle of interferometric detection of nanoparticles in a reflection-based geometry within an imaging chamber containing two interfaces representing a common sample configuration encountered in many flow cell or microfluidic chip designs. Arrows represent the different electric field contributions. (b) Image formation process in partially coherent interferometric systems for different light sources and degrees of spatial coherence. The image frames show a zoomed-in area of around $10 \times\SI{10}{\micro\metre}^2$ from total sample illuminated, and the contrast range is restricted to $\pm$ 0.07. \textbf{E$_{i}$:} incident electric field, \textbf{E$_{s}$:} scattering electric field, E$_{r1}$: reflection electric field from bottom glass/water interface, \textbf{E$_{r2}$}: reflection electric field from top water/glass interface, \textbf{BFP}: objective back focal plane,\textbf{ NA$_{obj}$}: numerical aperture of the objective, \textbf{r$_{obj}$}: aperture size at the back focal plane of the objective, \textbf{r$_{obj}$}: illumination beam size at the back focal plane of the objective,  \textbf{iPSF}: interferometric point spread function, \textbf{OTF}: optical transfer function, \textbf{k$_{NA}$}: spatial frequency corresponding to the numerical aperture of the detection objective.} 
      \label{fig:Intro}
  \end{figure}

The requirement for high photon flux sources is typically satisfied using high-power CW lasers, which are both spatially and temporally coherent illumination sources. However, achieving high sensitivity at high throughput with these light sources, defined here in terms of size of the field of view (FOV) per unit time, remains a significant challenge. On the one hand, coherent artifacts such as speckles and parasitic interferences severely degrade the image quality for wide-field imaging\cite{Shin2017EffectsMicroscopy}, and on the other hand, reducing these artifacts by turning the imaging system into a partially coherent one comes at the expense of a smaller FOV. To best capture these differences, Fig. \ref{fig:Intro} illustrates the image formation principle in a reflection-based interferometric microscope for three different scenarios: i) a coherent imaging system with a spatially coherent light source, ii) a partially coherent imaging system with a spatially coherent light source, and iii) a partially coherent imaging system with a spatially incoherent light source. The first case (Fig. \ref{fig:Intro}b, left column) is represented by Koehler illumination, where larger FOVs result from focussing the light source tightly into the BFP of the objective, essentially lowering the spatial frequency bandwidth, which in turn increases the spatial coherence of the imaging system; thus leading to coherent artifacts and lower spatial resolution, as represented in the optical transfer function (OTF). The second case (Fig. \ref{fig:Intro}b, middle column), represents a partially coherent microscope resulting from increasing the size of the illuminating beam at the BFP, which can be experimentally achieved by weakly focussing a laser into the BFP, or confocally illuminating the sample. Under this scenario, the spatial frequency bandwidth increases, thus lowering the spatial coherence and increasing the resolution of the imaging system. However, this reduces the FOV as shown in the middle column of Fig. \ref{fig:Intro}b. Solutions to extend the FOV exist, either in the form of rapid beam scanning of a weakly focussed beam with acousto-optic beam deflectors \cite{Ortega-Arroyo2012InterferometricMicroscopy}, raster scanning confocal detection \cite{Jacobsen2006InterferometricInterface, Kuppers2023ConfocalCells}, spinning disk confocal \cite{Hsiao2022SpinningCells}, or rotational integration of oblique scanning\cite{Liu2025UsingProtrusions}, yet come with drawbacks such as high peak intensities and limited scanning speeds that may ultimately restrict the throughput. The third case, showcases an alternative solution to this problem, which involves using a spatially incoherent illumination source \cite{Avci2015InterferometricDetection, Simmert2018LED-basedImaging, Yi2020InverseMicroscopy, Stollmann2023MolecularPlatform, Wu2025SimplifiedImaging}, such as multimode fiber-coupled light emitting diodes (LEDs), which not only delivers larger FOVs with minimal coherent artifacts, but also flat-top illumination profiles (Fig. \ref{fig:Intro}b right column).

Although illuminations with large spatial coherence degrade the image quality of label-free interference-based microscopy, and partial coherent imaging systems should be preferred for imaging weakly scattering objects, the question how one should tailor the degree of partial coherence with spatially incoherent light sources to maximise the signal-to-noise ratio (SNR) to quantify single particle signals in the least amount of time is lacking. This study addresses this gap and provides a solution applicable to metallic, dielectric, and biological nanoparticles. To do so, we developed a platform that allows simultaneous tuning and measurement of the degree of partial coherence. We specifically characterised the dependence of the signal contrasts and the background noise contributions to find the experimental parameters that not only optimise the SNR and acquisition throughput, but also reliably detect all NPs within the same focus position. We further demonstrated that such partially coherent systems are compatible with differential imaging modality, thus enabling sensitivities compatible for single protein detection but with orders of magnitude larger FOVs compared to state-of-the-art. The results from this work show that partially coherent systems provide a route to pushing the sensitivity and throughput of state-of-the-art interference-based label-free microscopes to new boundaries.

\section{Results and discussion}
\subsection{Working principle and experimental implementation}
\begin{figure}[h]
    \centering
    \includegraphics{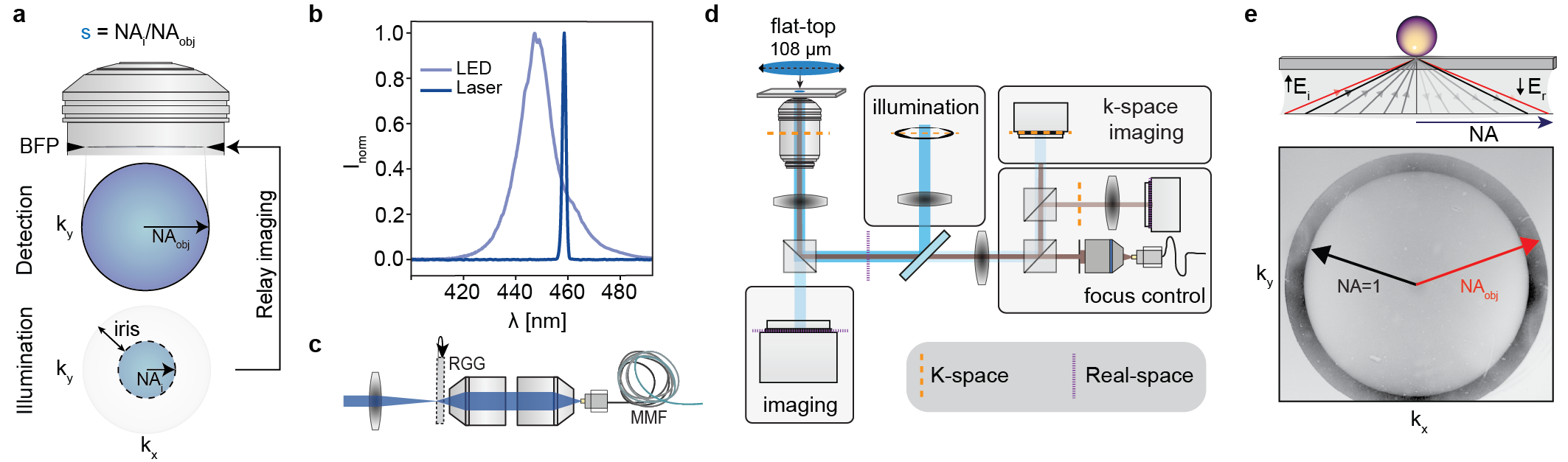}
    \caption{Working principle and experimental implementation.(a) Cartoon depicting the tuning of the degree of partial coherence. (b) Spectra of the two sources used for illumination: LED and laser diode. (c) Scheme for turning a laser into a spatially incoherent light source with a rotating diffuser and multimode fibre. (d) Experimental setup schematic with different lines indicating the locations of the real- and k-spaces. (e) Representative k-space image of a glass/air interface for measuring NA$_{i}$. Image taken for a fully open adjustable iris.}
    \label{fig:setup}
\end{figure}

In this work, we quantitatively assessed how spatial coherence influences the detection sensitivity at the single particle level. To achieve this, we incorporated a module that precisely controls and measures the spatial coherence into an existing reflection-based interferometric microscope. The degree of spatial coherence was quantified by the coherence parameter,$s$, as the ratio of the numerical aperture of illumination to that of the detection objective: $s$=NA$_{i}$/NA$_{obj}$.  Using this parameter we classified the imaging system into: spatially coherent for $s\rightarrow 0$, partially coherent for $0<s<1$, and incoherent for $s\geq 1$. To tune $s$ experimentally, we varied the size of an adjustable iris relay imaged to the backfocal plane (BFP) of the objective; effectively decoupling NA$_{i}$ from NA$_{obj}$  (Fig. \ref{fig:setup}a). To compare the performance of coherent versus partially coherent imaging systems, we used either a laser diode or a LED for illumination, with their respective spectra shown in Fig. \ref{fig:setup}b. In addition, light from the laser diode was turned into a partially coherent illumination source by focussing it onto a rotating ground glass diffuser (RGG) \cite{Dahlberg2017Step-by-stepDiffuser} before coupling into a multimode fibre (Fig. \ref{fig:setup}c). 

Fig \ref{fig:setup}d depicts the experimental setup comprising a custom-built common-path interferometric microscope operating in reflection mode featuring four different modules: imaging, focus stabilisation, light source input, and measurement of the NA$_{i}$. The focus control module together with the XY motorised sample stage enabled automation of the XYZ sample scanning assays with focus position stabilisation to within 10 nm. The NA$_{i}$ was tuned in the illumination module by adjusting the size of the iris, and subsequently measured in a separate k-space imaging channel. Fig. \ref{fig:setup}e shows a representative k-space image for a sample composed of immobilised nanoparticles at the glass-air interface. The observed bright ring corresponds to the total internally reflected incident angles of illumination (NA$_{i} \geq1$), with the inner and outer radii of the ring corresponding to NA=1 and NA$_{obj}$, respectively. Using the value of NA=1 from the glass-air as an internal reference\cite{Dai2005MeasuringMolecules}, we determined all input NA$_{i}$ and therefore $s$.

\subsection{Comparison between spatially incoherent light sources for a partially coherent microscope}
One of the main advantages of using LEDs in partially coherent imaging systems is the simultaneous reduction of speckle noise and access to large FOVs. However, their lower photon flux compared to lasers restricts the available photon budget for either sensitivity or throughput, but not both. To overcome this limitation, we converted a laser into a spatially incoherent light source, by increasing its spatial frequency bandwidth when illuminating through a RGG and coupling the transmitted light into a multimode fibre with more than 50$\%$ efficiency. This effectively suppresses coherent imaging artifacts associated to the laser via angular, spatial, and temporal domain averaging. 

\begin{figure}[H]
    \centering
    \includegraphics{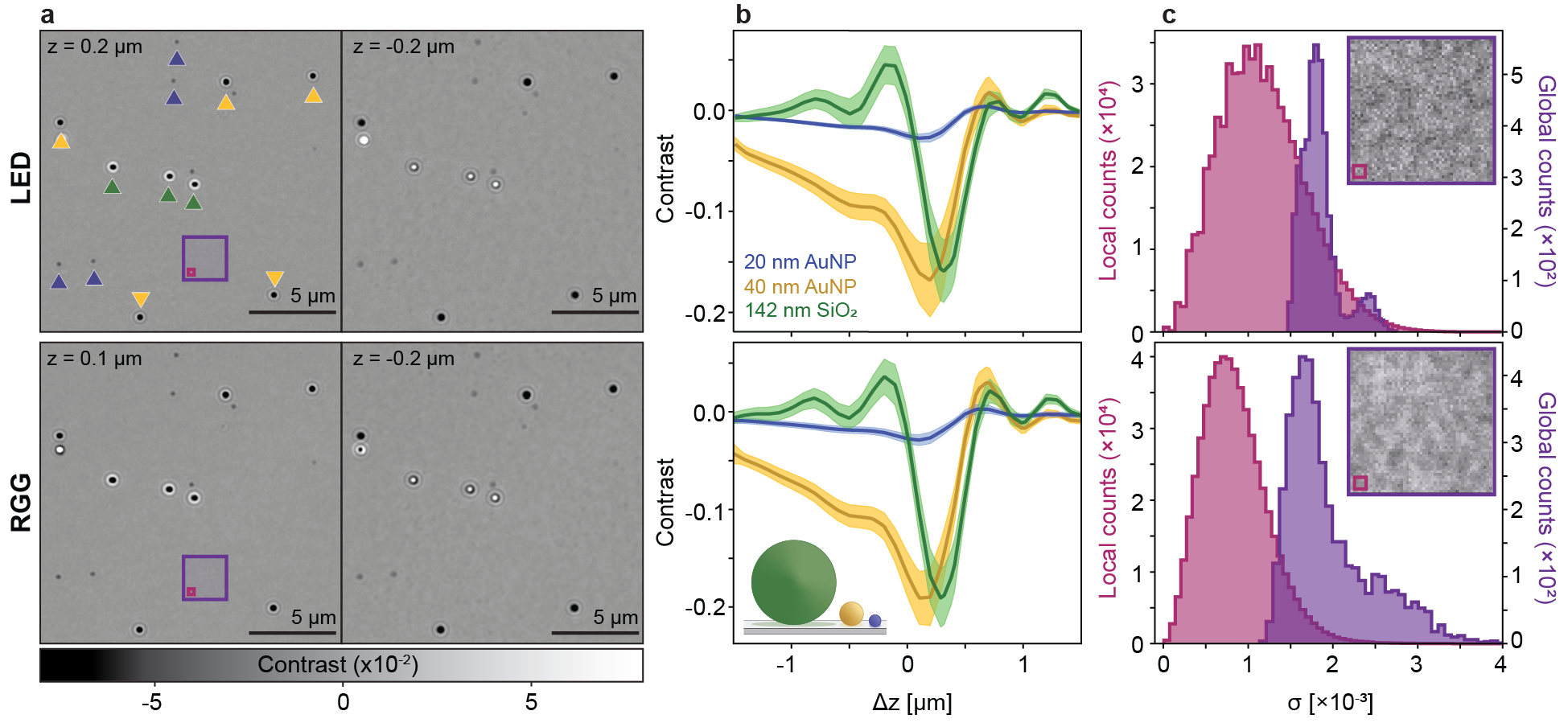}
    \caption{Effect of light source for delivering partial coherence. (a) Representative images of the same sample region area at two different focus positions illuminated with two different spatially incoherent sources. The polydisperse NP sample consists of 20 nm AuNP, 40 nm AuNP and 142 SiO$_{2}$ NP sample. (b) Corresponding distribution of particle contrasts as a function of focus position. Each colour represents a different particle population, with solid lines representing the mean ,and the shaded regions to $\pm$ one standard deviation. The number of detected nanoparticles consisted of 20nm AuNP (241), 40 nm AuNP (136),  and SiO$_{2}$NPs (31) for the measurement performed using the LED. For the RGG experiment the number of detected nanoparticles consisted in  20nm AuNP (285), 40 nm AuNP (105) and SiO$_{2}$NPs (51). (c) Local (pink) and global (purple) background noise distributions measured in standard deviations ($\sigma$) for the different light sources. The corner images depict an empty region of the sample, highlighted by the red box in (a), with the contrast adjusted to $\pm 0.01$ to emphasize the background noise. The areas within the image indicate the regions used to calculate global and local standard deviations for the histograms, with one standard deviation computed per region. The pink area includes 51$\times$51 pixels$^{2}$ and the purple area 3$\times$3 pixels$^{2}$, with one pixel corresponding to an area of 45$\times\SI{45}{\nano\metre}^2$.}
    \label{fig:LEDvsRGG}
\end{figure}

To validate the equivalence between the two light sources at the same fluences, irrespective of particle size and refractive index, we evaluated the particle contrast and image noise from a polydisperse sample containing 20 nm Au, 40 nm Au and 142 nm SiO$_2$ nanoparticles. To increase particle statistics the sample was raster-scanned over a total area of roughly $450 \times \SI{450}{\micro\metre}^2$ , corresponding to N = 10 FOVs of approximately $45 \times \SI{45}{\micro\metre}^2$. Notably, the sensor limited the size of the FOV recorded from a single image, given the nearly 4.5$\times$ larger illuminated area ( $\SI{9331}{\micro\metre}^2$). To ensure that the optimal signal contrast for each NP type was recorded, at each sample position the focus was scanned across 3 $\mu$m in 100 nm steps, denoted here as a defocus scan. Fig. \ref{fig:LEDvsRGG}a shows representative images of the polydisperse sample deposited on the glass coverslip at two different focus positions.

To characterise the particle contrast as a function of defocus, individual particles were localised and subsequently classified. Defocus scans of samples containing only one particle species at a time served as a reference for classification. Fig. \ref{fig:LEDvsRGG}b shows the average contrast curves  $\pm$ one standard deviation (shaded area) for each particle population. The shaded regions reflect the intrinsic size dispersion of each particle distribution. These scans showed a characteristic oscillatory behaviour between positive and negative contrast, pattern which was distinct for each of the three particle species. Furthermore, the maximum contrast magnitude for each particle type occurred at different defocus positions, consistent with contrast tuning with the Gouy phase \cite{OrtegaArroyo2014Label-freeProtein, Kukura2008ImagingDark} and the phase transfer function \cite{Zuo2017High-resolutionIllumination, Rosen2024Roadmapinvited}. Minor deviations between contrast curves from the two different light sources were attributed to slight spectral differences. Nonetheless, these defocus scans demonstrated the equivalence between both illumination schemes and the potential to use these scans as particle classifiers. 

To compare these two illumination schemes with respect to the background noise, we analysed two noise metrics corresponding to the local and global fluctuations within each image. Local noise fluctuations, quantified the shot noise within the image; whereas global noise fluctuations, predominantly measured the speckle and background roughness contributions. As a first step, we segmented all pixels within an image corresponding to the background, i.e. excluding those counted as particles. Local background noise was computed as the standard deviation within an 3$\times$3 background pixel area. The choice of an interrogation area significantly smaller than the diffraction limit, minimised any speckle or substrate roughness contributions. Global noise was calculated as the standard deviation within an 51$\times$51 background pixel area. Fig. \ref{fig:LEDvsRGG}c shows the distribution of local (pink) and global (purple) background noise from all frames under the two illumination schemes, with the global background noise at least two-fold higher than the local one. As expected from a shot-noise limited measurement, both partially coherent schemes showed comparable local background noise levels when illuminated at similar fluences. Similarly, speckle and substrate roughness contributions increased the noise level above shot-noise; with slightly higher values for the RGG-based illumination due to the presence of low spatial frequency components that had not been effectively suppressed during the integration time of the sensor.

In summary, for a partially coherent microscope, lasers combined with a RGG can perform just as well as LEDs and additionally deliver higher fluences. This enables imaging of larger fields of view at higher temporal resolutions, thereby increasing the overall throughput of the imaging system. Given the equivalence of the two light sources, all subsequent results were performed with a RGG-based laser as the illumination source.

\subsection{Effect of partial coherence on the signal to noise ratio for particle detection}
To determine how spatial coherence affects particle detection SNR, we repeated the defocus scan of the polydisperse sample under different coherence parameters $s$, by varying NA$_{i}$ but keeping the detection NA fixed and imaging the exact same sample area.  Fig. \ref{fig:zscan1}a shows representative zoomed-in regions containing all three NP species with their respective ensemble defocus scan contrast curves.  For these representative images, we chose the focus position at each parameter $s$ that maximized the contrast for the smallest particles (20 nm AuNP), as these were not the same for all particles or for different NA$_{i}$. Specifically, the focus position of the maximum negative contrast, indicated by the dotted vertical lines in Fig. \ref{fig:zscan1}a, shifted to higher defocus positions as the degree of partial coherence increased. Similarly, the amplitude of contrast oscillations in the contrast defocus curves defocus decayed with increasing $s$. To further validate our experimental data, we developed an imaging model for the detection of NPs as a function of defocus for partially coherent systems based on interferometric detection in reflection geometry(see \ref{sec:SImodel}). For all three NPs, model and experiment showed excellent agreement (Fig. \ref{fig:SImodel}). 

As expected from partially coherent imaging systems, as $s$ (NA$_{i}$) increased, the optical resolution also increased whilst the speckle contrast, PSF ringing and substrate roughness contributions, mostly composed of high spatial frequency components, decreased. In addition, the signal contrast for each particle increased as a function of $s$, before dropping slightly as the NA$_{i}$ approached the value of the refractive index of the solution. We attribute this latter drop in contrast to the increase in effective reflectivity associated with including total internal reflection contributions (see \ref{sec:SIreflectance}). 

To evaluate how the increase in particle contrast together with the reduction in background noise with increasing $s$ values translates to particle detection SNR, we first isolated each contribution individually. Fig. \ref{fig:zscan1}b shows the average contrast magnitude for all three NP species, showing at least a two-fold enhancement compared to the lowest $s$ value evaluated. Notably the smaller the particle, the higher the contrast enhancement and its occurrence at higher degrees of partial coherence.

For the noise component, we compared both the local and global background noise metrics (Fig. \ref{fig:zscan1}c). The local background noise remained constant as expected from illuminating at similar illumination fluences for the different partially coherent systems with values within the range of a shot-noise limited measurement based solely on camera counts. In contrast, the global noise level decreased approximately two-fold as a result of the reduction of coherent artifacts. We would like to emphasize that the reduction in noise was independent of NP sample type, and defocus position. 

We then computed the SNR as the ratio of the population average particle contrast magnitude to the global background noise (Fig. \ref{fig:zscan1}d). The overall trend showed that the SNR can be increased between two and four-fold within the range of partial coherence parameters tested with an optimal window within 0.7$<$NA$_{i}<$1.3. It should be emphasized that these enhancement factors underestimate the true enhancement relative to the coherent case, typically associated with widefield interferometric scattering microscopy (with $s$ $\ll$ 0.1). This is simply because coherent artifacts degraded the image so severely (Fig. \ref{fig:Intro}, representative image bottom left corner) that any quantitative particle characterisation was intractable.

\subsection{Partial coherent detection applied to biological nanoparticles}

\begin{figure}[H]
    \centering
    \includegraphics{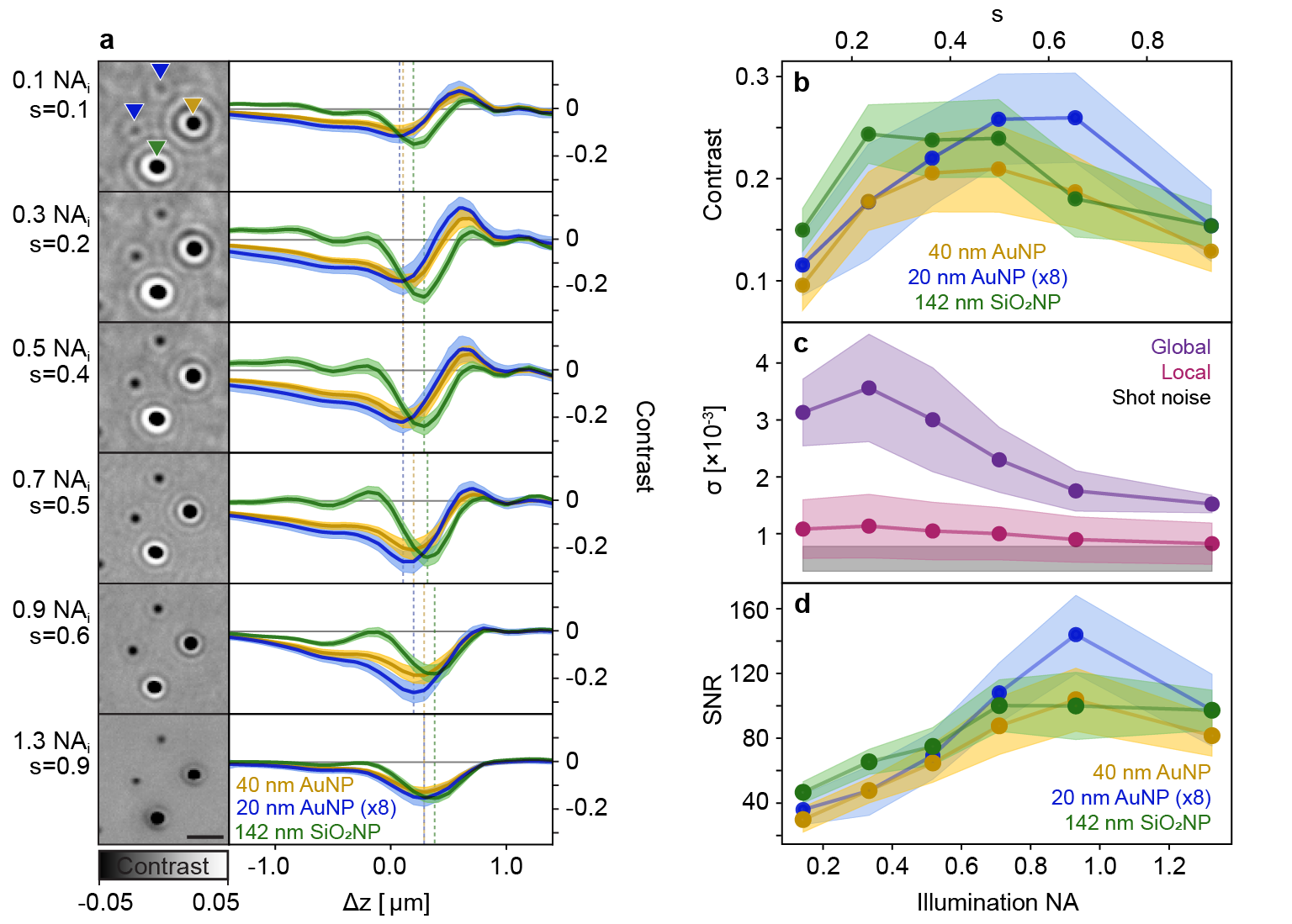}
    \caption{Effect of partial coherence on single NP detection. (a) Left: Inline holography images of two 20 nm AuNPs (blue triangle), a 40 nm AuNP (orange triangle) and a 142 nm $\mathrm{SiO}_2$ (green triangle). The images were acquired at the focus position where the smallest constituents, 20 nm AuNPs, exhibit the largest contrast, as indicated by the blue vertical markers in the right plots. Scale bars: $\SI{1}{\micro\metre}$. Right: Population-averaged contrast plotted against the z-position of the focus for the three particle species. Increasing from the smallest to the largest NA$_{i}$, the particle counts of each NP specie are: 20 nm AuNPs (420,428,404,424,249,291), 40 nm AuNPs (249,214,245,183,203,179), and SiO$_2$NPs (452,1302,1601,1409,1582,1828). (b) Population-averaged values of maximum absolute contrast  plotted against degree of spatial coherence. (c) Local and global background noise as a function of degree of spatial coherence. The black region shows the region of shot noise limit. (d) Population-averaged SNR as a function of the degree of spatial coherence, retrieved by dividing the population-averaged contrast shown in (b) by the global noise shown in (c).}
    \label{fig:zscan1}    
\end{figure}

\begin{figure}[h]
    \centering
    \includegraphics{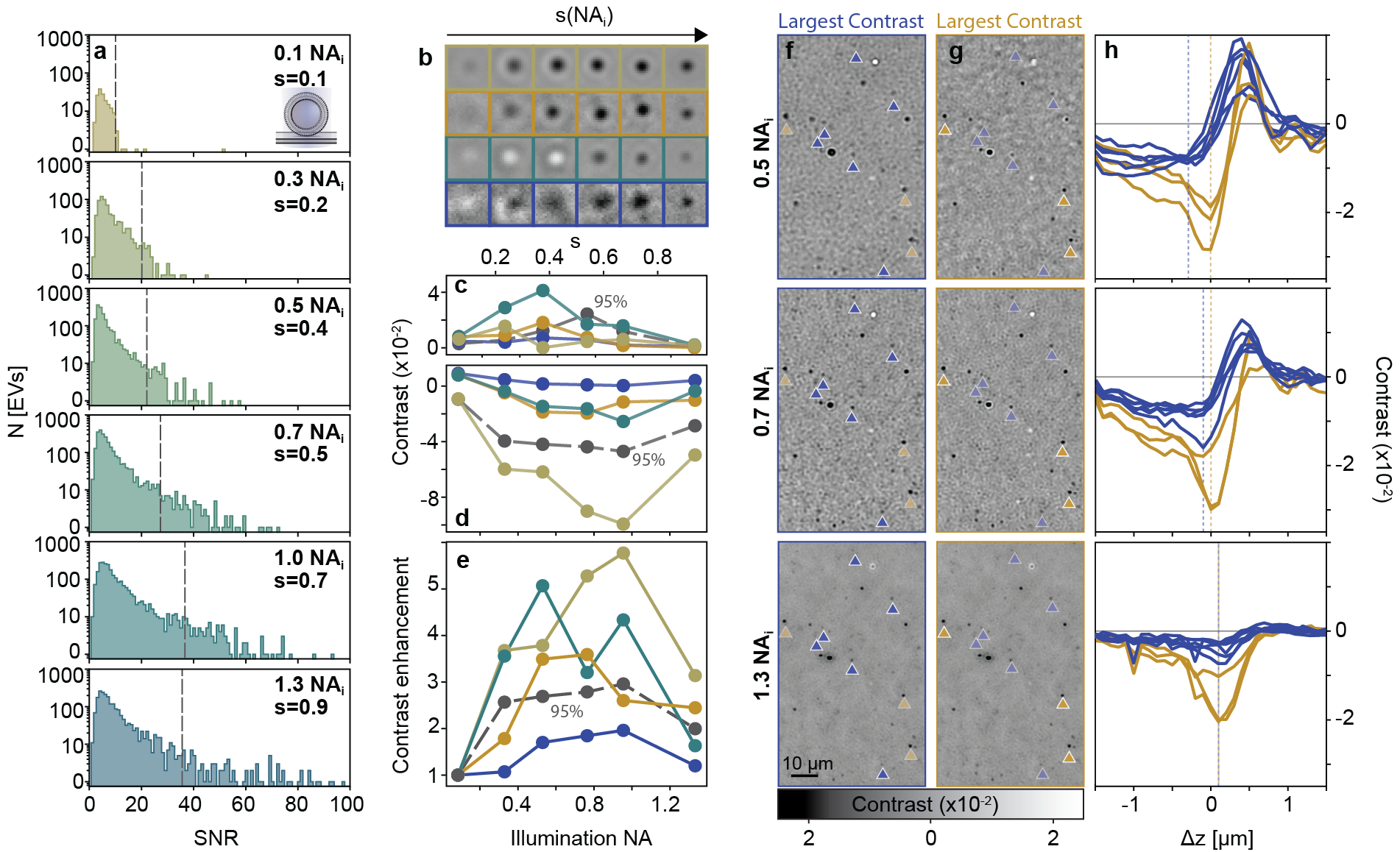}
    \caption{Effect of partial coherence on biological nanoparticle detection. (a)  Distribution of the maximum SNR of all detected EVs as a function of the degree of partial coherence. Increasing from the smallest to the largest NA$_{i}$, the counts of considered EVs are: 160, 683, 1746, 2501, 2491, 2098. (b-c) Representative zoomed-in images from the defocus scan at three selected coherence parameters. The axial position was chosen such that the contrast of the EVs marked by the blue arrow (b) and yellow arrow (c) is maximized respectively. (d) Contrast defocus curves Of the EVs marked in (b) and (c) at the respective NA$_{i}$.  The vertical dashed line indicate the axial position at which the images in (b) and (c) were acquired.}
  
    \label{fig:NAEV}
\end{figure}

Next, we repeated the defocus scans as a function of $s$ on a sample containing H358 cell culture-derived EVs (Fig. \ref{fig:SI_EVcharacterisation}) to determine whether similar SNR enhancements are expected with biological NPs. Here, the intrinsic size and refractive index heterogeneity of EVs make them an ideal system for observing general trends that extend to other biological NPs. Fig. \ref{fig:NAEV}a plots the SNR distribution of all single EVs detected for varying coherence parameters with the dashed dotted lines on each distribution indicating the 95th percentile. For the SNR calculation, we assumed that the optimal contrast had a negative value for all particles. Similarly to synthetic NP assays, increasing $s$ led to SNR enhancements for the EV sample, with the the maximum occurring in the range of 0.7$<$NA$_{i}<$1.3. From an ensemble perspective, an almost fourfold enhancement was observed, with the contrast enhancement contributing more than 60$\%$ to this increase (Fig \ref{fig:SI_EVcontrastDist}). Nevertheless, we must point out that this metric corresponded to a lower bound, as many EVs were not detected at lower $s$ values simply because their SNR fell below the detection threshold. 

To better estimate the range of SNR enhancement in this highly heterogeneous NP system, we examined how the contrast varies as a function of $s$ at the single-particle level; given that the noise contributions remain invariant across particle types. Fig. \ref{fig:NAEV}b shows four representative EVs, with each zoomed-in image along a row corresponding to the focus position that optimises the signal contrast for a given degree of partial coherence. In contrast to synthetic NPs (Fig. \ref{fig:zscan1}b), EVs displayed a more complex contrast dependence on $s$, including contrast inversions for some particles. To monitor this contrast inversion, Figs. \ref{fig:NAEV}c-d show the maximum positive and minimum negative contrast values for each EV, extracted from their respective contrast defocus curves (Fig. \ref{fig:SI_allsingleEVs}), with the dotted line marking the 95th percentile of the ensemble. Except for the EV with the largest contrast magnitude (olive line), all others underwent a signal contrast inversion - positive at low $s$ and negative at high $s$. This highlights the complex role both the Gouy phase and the phase transfer function play in modulating signal contrast at lower degrees of partial coherence. Finally, we estimated the enhancement as the absolute value of the optimal contrast at each $s$, normalised against the optimal contrast at the lowest tested $s$; with values ranging within two- and six-fold increase in the magnitude of the signal contrast. Once again, these values underestimate the overall enhancement relative to the coherent case, because of the highly detrimental coherent artifacts. Nevertheless, these observations indicate that optimal results were obtained for partially coherent systems within 0.7$<$NA$_{i}<$1.3; not only because of the overall contrast enhancement and accompanying reduction in background noise fluctuations, but also because of their consistent sign of the contrast signal; thus reducing ambiguity within the choice of focus position to optimally image at.

A crucial step in single-particle based sensing applications aimed at heterogeneous samples is the detection and subsequent characterisation of NP contrast signals, which are often only taken at a single focus position. However, as shown in Fig. \ref{fig:zscan1}b focus positions that maximise such, strongly depend on both the particle properties and the $s$ parameter. To illustrate how critical this scenario is at low $s$ imaging systems, Fig. \ref{fig:NAEV}f-g shows representative images of EVs immobilised on the surface taken at two different focus positions obtained from a defocus scan alongside the contrast defocus curves for a subset of EVs (Fig. \ref{fig:NAEV}h). Each focus position optimises the contrast of a particular subset of EVs indicated by blue (Fig. \ref{fig:NAEV}f) or orange (Fig. \ref{fig:NAEV}g) arrows. Note that for the measurements performed at NA$_{i}$=0.5, when the contrast magnitude of the subset marked in orange was maximised, the contrast of the subset marked in blue approached a zero crossing, causing the SNR of some of these EVs to fall below the detection threshold. This experiment serves as an example that interpreting particle contrasts from experiments performed at a single focus position warrants caution, as the measured contrast may not be representative for a given particle specie. However, as $s$ increased, this effect significantly reduced, as the distance between the two focus positions decreased and eventually converged to within 10 nm. 

These results lead to two different approaches for characterising heterogeneous NP samples that depend on the degree of partial coherence. If the degree of partial coherence is low, using a single focus position may severely underestimate both the count and contrast of entire subpopulations of NPs, instead, either the system PSF should be engineered \cite{Brooks2024PointCharacterization} to make it insensitive to defocus, or defocus scans should be measured; the latter offering potential advantages in the form of richer information content that can be used as particle classifiers \cite{Bon2015Three-dimensionalMicroscopy}. If the degree of partial coherence is high, a single focus position may suffice, as the information content about the particle properties within a contrast defocus decreases as all particle types tested tend to converge to a similar shape.

\subsection{Compatibility with differential imaging: single protein sensitivity}
\begin{figure}[H]
    \centering
    \includegraphics{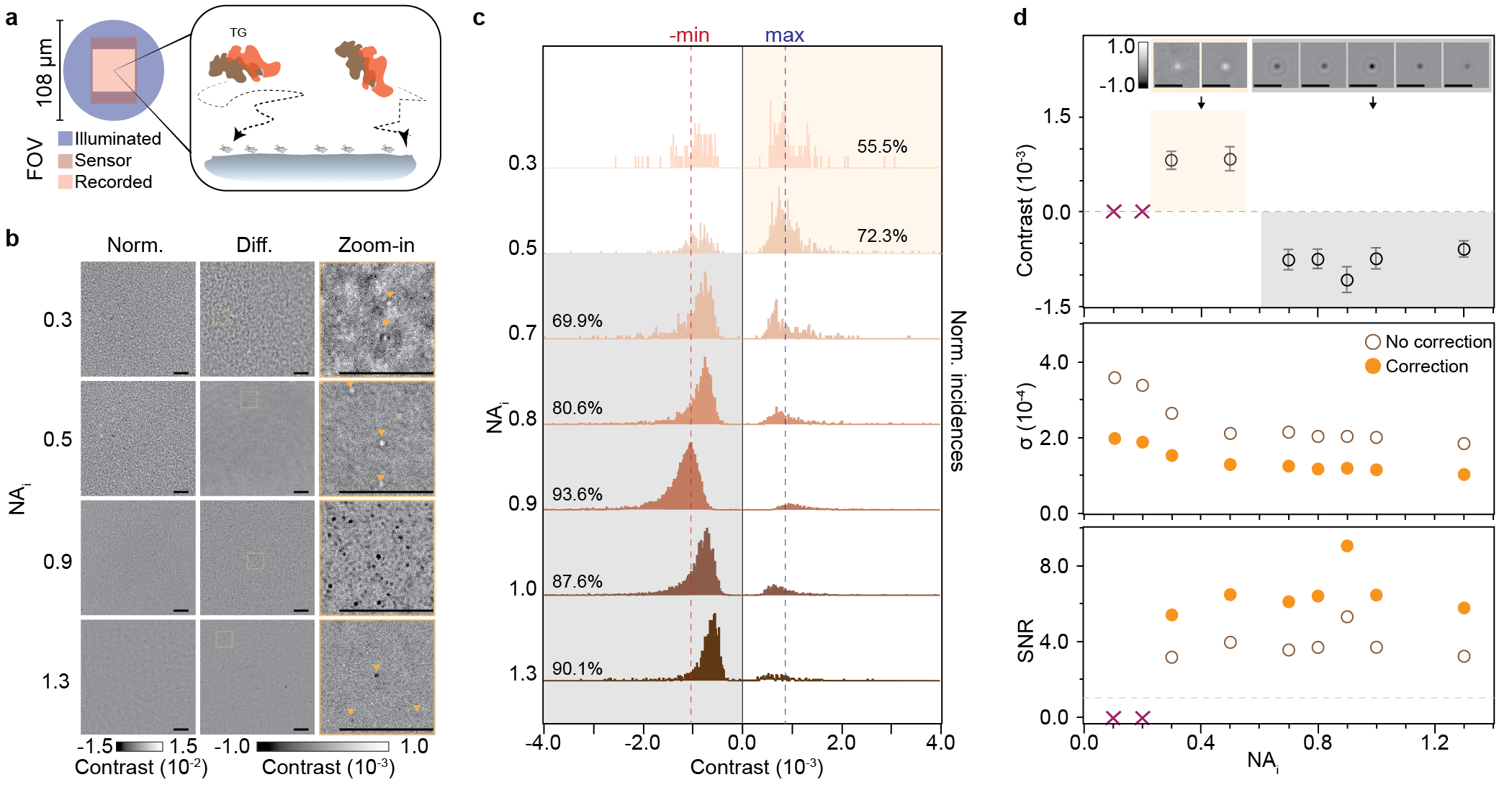}
    \caption{Partial coherence applied to differential imaging: single protein sensitivity. (a)  Schematic diagram for large FOV imaging of single  thyroglobulin (TG) proteins binding to a glass coverslip via differential imaging. (b) Representative images of single protein binding assays under different degrees of partial coherence. Orange box in the differential column represents the zoom-in region depicted in the right column.  Scale bars: 5 $\mu$m. (c) Contrast distribution of the detected particles as a function of degree of partial coherence. Shaded regions indicate the region of the distribution attributed to a binding event together with the percentage of the overall detected particles. The vertical dashed lines indicate the maximum and minimum contrast values of TG across all assays. (d) Corresponding contrast, noise and SNR of TG as a function of degree of partial coherence. Noise correction data involves applying a 2D spatial median filter of kernel size 55 pixels to reduce residual speckle and beam profile inhomogeneities in the differential images. Inset: ensemble averaged PSF. Scale bars: $\SI{1}{\micro\meter}$. Crosses indicate experimental conditions where it was not possible to detect TG.}
    \label{fig:TG}
\end{figure}

Finally, most interferometric-based microscopies leverage the intrinsic shot noise-limited nature of detection in the form of differential imaging, whereby a continuously updated background is subtracted from an ongoing set of images, combined with frame averaging to drastically improve the sensitivity limits of a single shot acquisition. To confirm that tuning the partial coherence also enhances the SNR under this imaging modality, we chose a test assay ubiquitiously encountered in label-free protein detection and mass photometry: quantification of the non-specific binding of single thyroglobulin proteins (TG, 660 kDa, dimer) onto a glass coverslip (Fig. \ref{fig:TG}a). We specifically performed this assay on the same FOV at different degrees of partial coherence, NA$_{i}$ =$\{$0.1,1.3$\}$, keeping the camera counts the same; thereby ensuring equal photon statistics between experiments. By observing the same FOV, we allowed the imaging system to mechanically relax and thus minimised any significant lateral sample drifts. For each imaging condition, the focus position was optimised to yield the maximum absolute signal contrast of TG. 

Figure \ref{fig:TG}b shows representative normalised and differential images of the glass surface with TG binding highlighted in the zoom-in regions. Upon changing from lower degrees of partial coherence to higher ones, a clear contrast signal inversion and reduction in background noise were observed, in agreement with the results from the extracellular vesicles assays. All TG binding and unbinding events were subsequently localised and their respective normalised contrast distribution and fraction of binding events to total particles plotted in Fig. \ref{fig:TG}c. No particles were detected for NA$_{i}< 0.3$, as they fell below the detection threshold by a combination of lower signal contrast and higher background noise levels. Differences in the fraction of binding events reflected the increase in false positives attributed to the inclusion of residual speckle contributions with high structure similarity to TG PSFs. Figure \ref{fig:TG}d summarises the signal contrast, background noise level and resulting SNR as a function of the degree of partial coherence for the ensemble of TGs measured, with the inset showing the ensemble-averaged PSFs at each measurement condition. Overall, these data show a trend similar to synthetic and biological NPs, with the lowest SNR at low NA$_{i}$, a contrast inversion occurring between NA$_{i}$= 0.5 and 0.7, a maximum contrast around NA$_{i}$= 0.9, followed by a decrease in signal contrast and a monotonic decay in background noise as NA$_{i}$ approaches 1.3, where total internal reflection contributions become significant. 

These results demonstrate that partial coherent systems are also compatible for differential-based imaging and offer similar SNR tuning as for larger NP systems. Despite the achieved sensitivity is lower compared to specialised label-free protein detection systems, our FOVs are orders of magnitude higher, and involve the use of mechanically oscillating RGG systems and cheap laser diodes with poor beam quality.  We believe that the sensitivity and throughput can be further improved by engineering the illumination beam profile (see Fig. \ref{fig:SI_BeamEng}), increasing the camera frame rate, the speed of the rotating diffuser; the latter two reduce the effect of mechanical drift or time-varying artifacts from the laser and RGG system. As a more suitable solution for differential imaging applications,  we propose the use of high intensity LED systems, as LEDs remove the need for any rotating mechanical elements, offering better stability, suppression of residual time-varying speckles from imperfect RGG synchronisation and laser mode hopping, and additional lowering of the temporal coherence\cite{Stollmann2023MolecularPlatform} (Peer Review).

\section{Conclusion}
In this work, we showcased a platform that simultaneously tunes and measures the degree of partial coherence with the aim to quantify how this influences the detection sensitivity of single nanoparticles. We further recapitulated the main experimental findings with an imaging model for partially coherent systems.  By characterising the particle signal at different focus positions and the background noise we demonstrated that a diode laser can achieve performance similar to that of an LED, yet with the advantage of a higher available photon flux. Our results on tuning the spatial coherence to intermediate $s$ values show that the SNR for the detection of metallic, dielectric and biological nanoparticles, the latter of which includes single proteins, can be enhanced compared with the coherent case due to a synergistic combination of background noise reduction and signal contrast enhancement.

With the defocus scans we further showed the pivotal role the degree of partial coherence plays on modulating the signal contrast response for different particle types. In the case of imaging systems with low $s$ parameter, the fact that there is no unique focus position that optimises the contrast for all particle types within heterogeneous NP samples, can lead to entire subsets of NPs going undetected when optimizing the contrast for a specific NP population. This highlights the importance of acquiring defocus scans in these imaging systems because they provide valuable information that can be exploited for sizing\cite{Zambochova2023AxialSize}, classification \cite{Bon2015Three-dimensionalMicroscopy}, or sample tilt compensation\cite{Aygun2019Label-freeMicroscopy}. One way to retrieve the axial information, besides time-consuming defocus scans, is to retrieve the phase and performing digital propagation, for instance, by solving the Transport-of-Intensity equation\cite{TEAGUE1982IrradiancePhase} or performing these measurements with an off-axis holography configuration. Alternatively, if throughput and sensitivity is paramount, then imaging systems with high degrees of partial coherence should be preferred.

Lastly, we have demonstrated that partial coherence imaging is compatible for differential-based detection, thereby promising to increase the throughput of assays, in terms of total FOV imaged, that rely on the quantification of spatially varying heterogeneous signals that fall below the signals levels of the static background; which can either be in the form of proteins, nucleic acids, lipid nanoparticles, or different charge states. All in all, we believe our work paves the way towards democratizing how inline holographic approaches based on interferometric detection can deliver both high sensitivity and throughput without the need of beam scanning solutions. %

\section{Materials and Methods}
\subsection{Microscope} 
The custom-built partially coherent digital holographic optical system was based on a common-path microscope operating in reflection, whereby illumination and imaging arms were separated by a single 50:50 beamsplitter plate (BSW27, Thorlabs). Partially coherent illumination was achieved by two approaches: Focussing a 462 nm laser beam (LDM-465-3000-C, Lasertack) on a rotating ground glass (RGG) diffuser (DG20-1500, Thorlabs) or using a 455 nm LED (M455F3, Thorlabs). For the first option, the laser was coupled out from a single-mode fiber (P1-460A-FC-2, Thorlabs) by a 0.1 NA objective (Olympus Uplan FLN). A plano-convex lens (LA1986, Thorlabs) focused the light on the RGG, which was driven at 600 rpm by a stepper motor (42BYG Stepper Motor, Makeblock). After the RGG, a 0.4 NA objective (olympus PlanN) collected the diverging beam. The laser beam with reduced coherence was coupled into a multi-mode fiber (FT600EMT, Thorlabs) by a 0.3 NA objective (Olympus Uplan FLN). Both options for partially coherent light sources could be coupled into the inline holography system by a 0.25 NA aspheric lens (C220TMD-A, Thorlabs). A relay system composed of two plano-convex lenses (LA1131 and LA1509-A, Thorlabs) allowed access to the back focal plane, in which an adjustable iris was placed to control the NA$_{i}$. The image plane was relay-imaged onto the sample plane via a 3:4 imaging system composed of two plano-convex lenses (AC508-400-A-ML and AC508-300-A, Thorlabs) and a 1.42 oil immersion objective (UPLXAPO 60X, Olympus). The flat-top illumination measured a diameter of $\SI{108}{\micro\meter}$. The same objective collected light from the sample. The 50:50 beamsplitter guides a part of the collected light to a 2:1 relay system composed of two plano-convex lenses (AC508-100-A-ML and AC508-200-A-ML, Thorlabs) and finally onto a CMOS camera ( $\SI{9}{\micro\metre}$, BFS-U3-17S7M-C USB 3.1 Blackfly S, Teledyne). The sensor area was around half the size of the image plane at that position, which restricted the detected FOV to $72\times\SI{50}{\micro\meter}$. This imaging system resulted in a $200\times$ magnification. The sample focus position was encoded and stabilised using the back reflection from a 670 nm beam (CPS670F, Thorlabs) confocally illuminating the sample. Specifically, the diameter of the reflected beam was used as a feedback parameter in a proportional–integral–derivative loop, making it insensitive to beam pointing instabilities. The sample was mounted on a motorised XY microstage (Mad City Labs) equipped with linear encoders, and a Z nanopositioner stage (Nano-Z200, Mad City Labs). 

\subsection{Optical imaging}
During acquisition, a field of view of $46\times\SI{46}{\micro\meter}^2$ corresponding to an area of $1024\times1024$ camera pixels$^2$ was recorded with an exposure time of 19.6 ms and a fixed frame rate of 50 Hz. To minimise data load and increase the signal-to-noise ratio, data were saved in the form of 20 time-averaged frames, leading to an effective time resolution of 2.5 Hz. The rotation speed of the diffuser was set to 600 rpm and synchronised with the camera frame rate, such that each effective time-averaged frame would include the average of four revolutions. For all synthetic and biological NP experiments, we measured a power at the sample of approximately 6.5 mW, equivalent to an irradiance of 0.7 $\mu$W/$\mu$m$^{2}$.

\subsection{Sample preparation}
For the experiments with synthetic NPs we used  142 nm SiO$_2$ NPs (SiO$_2$-R-L3205-23/1, Microparticles GmbH), 40 nm AuNP (AuXR40, nanoComposix), and 20 nm AuNP (EM.STP20, BBI Solutions). All nanoparticles were suspended in deionized water to a concentration of 8 pM for SiO$_2$ and 40 nm AuNP, and 16 pM for 20 nm AuNPs. Before introducing the NP sample onto the glass surface, each glass coverslip was cleaned with isopropanol and rinsed with deionized water. To locate the approximate focal position, 50 $\mu$L of deionized water was first deposited on the coverslip. NPs were then sequentially added by pipetting 1 $\mu$L of each stock solution onto the coverslip. After each addition, 5–10 $\mu$L of phosphate-buffered saline was introduced to favour non-specific binding of the NPs to the glass surface due to reduction of the Debye screening length. The coverslip was then rinsed with deionized water to remove excess particles. Before starting the measurements, an additional 50 $\mu$L of deionized water was added to prevent drying during data acquisition.

\subsection{EV isolation and characterisation}
The human lung cancer cell line H358 was purchased from the American Type Culture Collection (ATCC: CRL-5807). Cells were cultured in RPMI (ATCC formulation, Gibco A01491) supplemented with 10$\%$ fetal bovine serum (FBS, Gibco 10270-106), 1$\%$ Pen/Strep (Gibco 15140-122) at 37$^\circ$C in 5$\%$ CO$_{2}$. For EV isolation, cells were first detached with 0.25$\%$ Trypsin-EDTA (Gibco, 25200-056), centrifuged at 700$\times$g for 7 min, and the cell pellet washed with PBS (Gibco, 10010-015). Next 12$\times$T150 flasks were each plated with 2.5M H358 cells in 20 ml RPMI including 10$\%$ EV depleted FBS (Gibco, A27208-01) and 1$\% $Pen/Strep. After culturing for 72h to 60-70\% confluency, the supernatant was collected and centrifuged at 1,500$\times$g for 10 min, then at 10,000$\times$g for 10 min at 4$^\circ$C to remove floating cells or large debris. The supernatant was concentrated using an Amicon Ultra-15 centrifugal filter (MWCO = 50 kDa, Merck UFC905096) at 5000$\times$g for 30 min at 4$^\circ$C. The concentrated sample was then purified via size-exclusion chromatography column according to the manufacturer's specifications (Izon, qEV1 70nm). Namely for each 1 mL of isolated EV sample, 10 mL of PBS were added as an elution volume, from which the first 4.7 mL were discarded and the following 4 mL were collected as the EV fraction. The EV fraction was concentrated with the Amicon Ultra-15 Centrifugal filters and afterwards supplemented with 1$\times$ protease inhibitor cocktail (Thermo Scientific, 87786) before storing at -80$^\circ$C until further use.

EVs were lysed in 10$\times$ RIPA lysis buffer (Merck 20-188) for Western Blot analysis to confirm the characteristic EV biomarkers (CD9, CD63, and TSG101) and degree of purity using a non-EV marker(GRP94). The blots were probed with flowing primary antibodies: anti-CD9 (1:500 dilution, Thermo, 10626D), anti-CD63 (1:1000 dilution, Boster, M01080-1), anti-TSG101 (1:1000 dilution, Biorbyt, ORB1564135), and anti-GRP94 (1:1000 dilution, FineTest, FNab03665). Chemiluminescence was detected using an iBright CL1500 system (Thermo Scientific A44240) using SuperSignal West Pico Plus Chemiluminescence Substrate (Thermo Scientific 34277) and SuperSignal West Atto Ultimate Sensitivity Substrate (Thermo Scientific A38555). The concentration and mean size of the EVs was determined by nanoparticle tracking analysis using Zetaview (Particle Metrix), and found to be 2.8$\times10^{10}$ particles/mL and 124.4 nm, respectively.

\subsection{Image processing}
Each acquired frame was normalised by the median pixel value to correct for shot-to-shot power fluctuations. Sample-independent static contributions were consequently removed by flat-fielding. For this, the median image was computed from 16 different lateral positions at the same focus. Each frame was then divided by the median image computed for the given focus.  

\subsection{Particle localisation}
The first step in particle localisation consisted of creating SNR-enhanced images from the normalised flat-field images. For this, the normalised images were binned $2\times2$. Then, the root mean squared of the background was computed by only including pixels with values smaller than three times the global standard deviation. The standard deviation was estimated from the median absolute deviation. An SNR-enhanced image was created by dividing the binned images by the RMS of the respective background pixels.

All pixels with absolute values larger than 0.2 in the SNR-enhanced image were used as initial guesses for particle localisation, if they additionally fulfilled the requirement of being the local extrema within a $9\times9$ pixel window. These guesses were verified using Trackpy 0.6.4 and a radial symmetry fit. In the next step, Trackpy 0.6.4 was used to link the particles in the different axial positions. The particle guesses were further considered under the condition that they could be linked along a single trajectory over a distance of $\SI{1.5}{\micro\meter}$ under a memory of zero and a search range of 1.75.

\subsection{Label-free detection of non-specific binding of TG}
TG (Merck, T1001) was resuspended at 1 mg/ml in  water and passed through a 10000 kDa MWCO filter to remove aggregates. Experiments were performed in a microwell, whereby a solution of 10 nM of TG in PBS was injected prior to imaging. During acquisition of TG data, a field of view of $46\times\SI{46}{\micro\meter}^2$ corresponding to an area of $1024\times1024$ camera pixels was recorded with an exposure time of 4.62 ms and a fixed frame rate of 200 Hz. To minimise data load and increase the signal-to-noise ratio, data were saved in the form of 20 time-averaged frames, leading to an effective time resolution of 10 Hz, which was synchronised to the rotation speed of the diffuser to equal a single revolution. For all TG experiments we recorded 500 averaged-frames (50 s) with a power at the sample of approximately 26 mW, equivalent to an irradiance of 2.8 $\mu$W/$\mu$m$^{2}$.

For differential imaging, we computed the rolling differential window average ($\Delta_{i}$) for the i-th frame, $I_{i}$, as: 
\begin{equation*}
    \Delta_{i}=\frac{\sum^{N-1}_{j=0}I(i+j)-\sum^{N-1}_{j=0}I(i-N+j)}{N}
\end{equation*}
with N=50 representing the number of frames averaged. In total, each image in $\Delta_{i}$ corresponds to effectively averaging 1000 raw camera frames (effective frame rate 0.2 Hz). For detection of single binding and unbinding events, no further image processing other than a flatfield correction and a 2D spatial median filter of with kernel size M=55 pixels were used to remove illumination inhomogeneities caused by residual speckle, changes in laser mode, and beam-pointing instabilities. Only detection events with track lengths $\geq$ 30 times points were considered for further analysis.

\begin{acknowledgement}
The authors acknowledge the following funding sources: Swiss National Science Foundation grant 207485 (C.L., S.S., J.O.A., R.Q.); Bio-Engineering Systems for Therapeutics (BEST) postdoc programme established between ETH Zurich and F.Hoffmann-La Roche Ltd. (A.S.); French Investissements d’Avenir program grant 17-EURE-0002 (G.CdF) ; French National Agency grant ANR-24-CE42-5810 (G.CdF).\\
\end{acknowledgement}

\section*{Author Contributions}
Conceptualization: J.O.A,. Funding: R.Q., J.O.A. Methodology: C.L., J.O.A. Simulations: G.CdF. Experiments: C.L., A.S., S.S. Software: C.L., A.S., G.CdF., J.O.A. Formal analysis: C.L., A.S., G.CdF, J.O.A. Visualization: C.L., A.S., G.CdF, J.O.A. Supervision: R.Q., J.O.A. Writing original draft: C.L., J.O.A. Writing, reviewing and editing: C.L., A.S., S.S., G.CdF, R.Q., J.O.A.
\begin{suppinfo}
Model for partially coherent interferometric detection, origin of the contrast enhancement for partially coherent imaging systems, H358 EV characterisation, distribution of maximal contrast of all EVs detected as a function of degree of partial coherence, contrast defocus curves as function of degree of partial coherence for four representative EVs, illumination beam engineering enhances the signal contrast.
\end{suppinfo}

\bibliography{references,refGCF}
\makeatletter
\renewcommand \thesection{S\@arabic\c@section}
\renewcommand\thetable{S\@arabic\c@table}
\renewcommand \thefigure{S\@arabic\c@figure}
\makeatother

\section*{Supplementary information}
\setcounter{section}{0}
\setcounter{figure}{0}

\section{Model for partially coherent interferometric detection}
\label{sec:SImodel}
To gain better insight into our system we developed an imaging model based on partial coherence that recapitulates the experimental contrast defocus curves shown in Fig \ref{fig:zscan1}. This model builds on existing results from literature, and is divided into three sections,  whereby we specifically mention the approximations that can be made in the current configuration: i) the excitation and reflected field considering incoherent K{\"o}hler illumination, ii) dipolar scattering in the objective and iii) optical path aberration appearing in defocused interferometric microscopy.

\paragraph{Incident and reflected fields}
We follow ref. \cite{Unlu:2016} and model the K{\"o}hler illumination considering incoherent plane waves impinging on the glass coverslip interface at angles $(\theta_m,\phi_m)$ that span the numerical aperture of an oil immersion objective. The objective images the focus ($z=0$) given a specified glass substrate (optical index $n_{g}$, thickness $t_{g}$) and oil immersion film (optical index $n_{i}$, thickness $t_{i}$),  see Fig. \ref{fig:Kohler}. Deviations from optimal conditions will be introduced later using an aberration phase shift. We denote $r_{s}, r_{p}$ ($t_{s}, t_{p}$) as the the coverslip Fresnel coefficient of reflection (transmission) for the TE and TM polarisations, respectively. Note that the coverslip Fresnel coefficients are approximated for small oil/glass optical contrast, and the phase shift originating from the oil and glass thickness will be included in the aberration phase shift \cite{Haeberle:03}. Using the angular spectrum representation \cite{Novotny-Hecht:2012} we express the incident transmitted field that excites the nanoparticle as:

\begin{figure}[H]
\centering
\includegraphics[width=7.5cm]{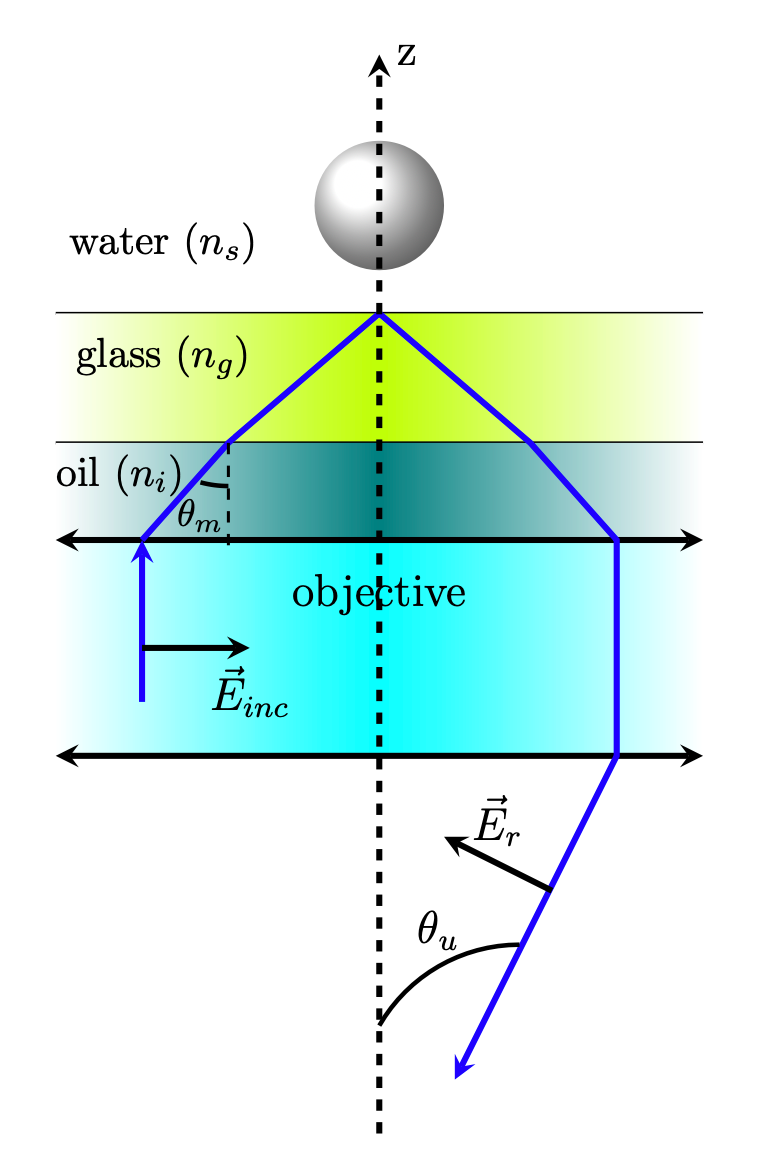}
\caption{Plane wave expansion model for incoherent K{\"o}hler illumination. Each plane wave is characterised by an incident angle $\theta_m$, which reflect on the coverslip and is finally detected at the image plane of the objective at the output angle $\theta_u$.}  
\label{fig:Kohler}
\end{figure}

\begin{eqnarray}
&&\vec E_{t}(O)=A_1E_0 \sin \theta_m \sqrt{\cos\theta_m} 
\begin{pmatrix}
t_p\cos^2 \phi_m \cos \theta_s+t_s\sin^2\phi_m \\
(t_p\cos \theta_s-t_s) \cos\phi_m \sin\phi_m \\
-t_p\cos \phi_m \sin\theta_s 
\end{pmatrix}
\label{eq:Et}
\\
&&\cos\theta_s=\sqrt{1-\frac{n_{i}^2}{n_s^2}\sin^2\theta_m}
\nonumber
\end{eqnarray}
In the following, we note $w_s=n_sk_0\cos \theta_s$ the vertical component of the wavector in water with the angle in water following the Snell-Descartes law $n_s \sin \theta_s=n_i \sin \theta_m$. Similar definitions occurs in all media. 
The coverslip Fresnel coefficients are approximated for small oil/glass optical contrast. Indeed, one can then write 
\begin{eqnarray*}
t^{slab}=\frac{t_{o/g}t_{g/w}e^{iw_gt_g}}{1+r_{o/g}r_{g/w}e^{2iw_gt_g}}e^{i(w_iz_i-w_sz_s)}\approx t_{o/g}t_{g/w}e^{i\Psi_t}
\end{eqnarray*}
where $\Psi_t$ is a phase shift that will be rewritten as an aberration optical path to describe the deviation from the design configuration \cite{Haeberle:03}. So that $t_{s/p}=t_{o/g}^{s/p}t_{g/w}^{s/p}$ has to be used in the above expressions.

In adddition, the reflected field imaged at the position of the detector writes:
\begin{eqnarray}
\nonumber
\vec E_{r}(O')&\approx&\frac{A_2}{M^2}E_0 \sin \theta_m\cos\theta_m 
\begin{pmatrix}
-r_p\cos^2 \phi_m +r_s\sin^2\phi_m \\
-(r_p +r_s) \cos\phi_m \sin\phi_m \\
0
\end{pmatrix}
\end{eqnarray}
Again, the Fresnel coefficient are approximated for small oil/glass optical contrast.
\begin{eqnarray*}
r^{slab}=\frac{r_{o/g}+r_{g/w}e^{2iw_gt_g}}{1+r_{g/w}e^{2iw_gt_g}}e^{-2iw_it_g}\approx   r_{g/w}e^{i\Psi_r}
\end{eqnarray*}
$\Psi_r$ is a phase shift that will included also in the an aberration optical path. $r_{s/p}=r_{g/w}^{s/p}$ has to be used.

In the above equations, focusing parameters associated to lens 1 and lens 2 are written as: $A_1=ik_0 f_1e^{-in_{i0}k_0 f_1}/2\pi$ and $A_2=ik_0 f_2e^{-ik_0 f_2}/2\pi$. In this description, we have used the approximations $\sin\theta_u  =\sin \theta/M \ll1$ and $\cos \theta_u \approx 1$ based on the large magnification $M$ of the microscope.  

\paragraph{Scattered field}
We approximate the scattered field from the NP as that of an induced dipole; namely: 
\begin{eqnarray}
\vec p=\alpha_p \vec E_t(O) \;; \alpha_p=4\pi R^3 \epsilon_p \frac{\epsilon_p-\epsilon_s}{\epsilon_p+2\epsilon_s}.
\end{eqnarray}

The dipolar electric field radiated by the dipolar source is determined from the angular representation $\vec E_o$. For a particle far from the glass/water interface, one can use the approximation \cite{Aguet:09,Dong:21} $\vec E_o=\left[(\vec p \cdot \vec e_\parallel)\vec e_\parallel+ (\vec p \cdot \vec e_\perp)\vec e_\perp\right]$ with $\vec e_{\parallel,\perp}$ TE/TM polarized unit vectors. However, for a particle on the substrate, the evanescent coupling cannot be neglected and the exact angular representation has to be considered \cite{Unlu:2016,Sentenac:19}.
We obtain at the detector position $O'=(0,0,0)$ in the image plane, again considering approximation for large microscope magnification $M$: 
\begin{eqnarray}
\nonumber
&&\vec E_{s}\approx I_{0}\left(p_x \vec e_x+p_y \vec e_y\right) \\
&&I_{0}\approx \frac{in_i^{3/2}k_0^3}{n_s M^2} \frac{A_2}{2A_1}\int_0^{\theta_{det}} \left(\tau_p +\frac{k_s}{w_s}\tau_s \right)  (\cos\theta)^{\frac{3}{2}} \sin \theta  \mathrm d\theta
\nonumber
\end{eqnarray} 
with $\tau_{s,p}^{slab}$ the Fresnel coefficients of transmission of the coverslip from water to oil (again approximated for small oil glass optical contrast). As an approximation, we assume a point-like detection at $O'$ and thus consider the scattering contribution of vertical dipole negligible for large magnification \cite{Novotny-Hecht:2012}. However, for a fully comprehensive model, the signal should be integrated over the sensor area, as the field scattered by a vertical dipole would also contribute to the detected signal (see also the discussion in \S \ref{sect:DipoleOrient}).
\paragraph{Aberration optical path}
\label{sect:Phase}
Under ideal conditions, oil immersion objectives are designed to focus at the glass/water interface when the specific glass substrate (optical index $n_{g0}$, thickness $t_{g0}$) and oil immersion film (optical index $n_{i0}$, thickness $t_{i0}$) are met. However, in practice, the glass substrate can have small deviations in $n_{g}$ and $t_{g}$; thus, the oil immersion layer has to be adapted accordingly (optical index $n_{i}$ , thickness $t_{i}$). Moreover spherical aberration affects the position of the focus. To account for spherical aberration, we apply a correction factor $\Delta_f$ between the movement of the focal plane $z_f$ and the movement of the objective $z_{exp}$:  $z_f=\Delta_f z_{exp}$ with \cite{Diel:20} 
$$\Delta_f=\frac{\tan\left(\sin^{-1}(0.5\mathrm{NA_{obj}}/n_i)\right)}{\tan\left(\sin^{-1}(0.5\mathrm{NA_{obj}}/n_s)\right)}$$

\begin{figure}[H]
\centering
\includegraphics[width=10cm]{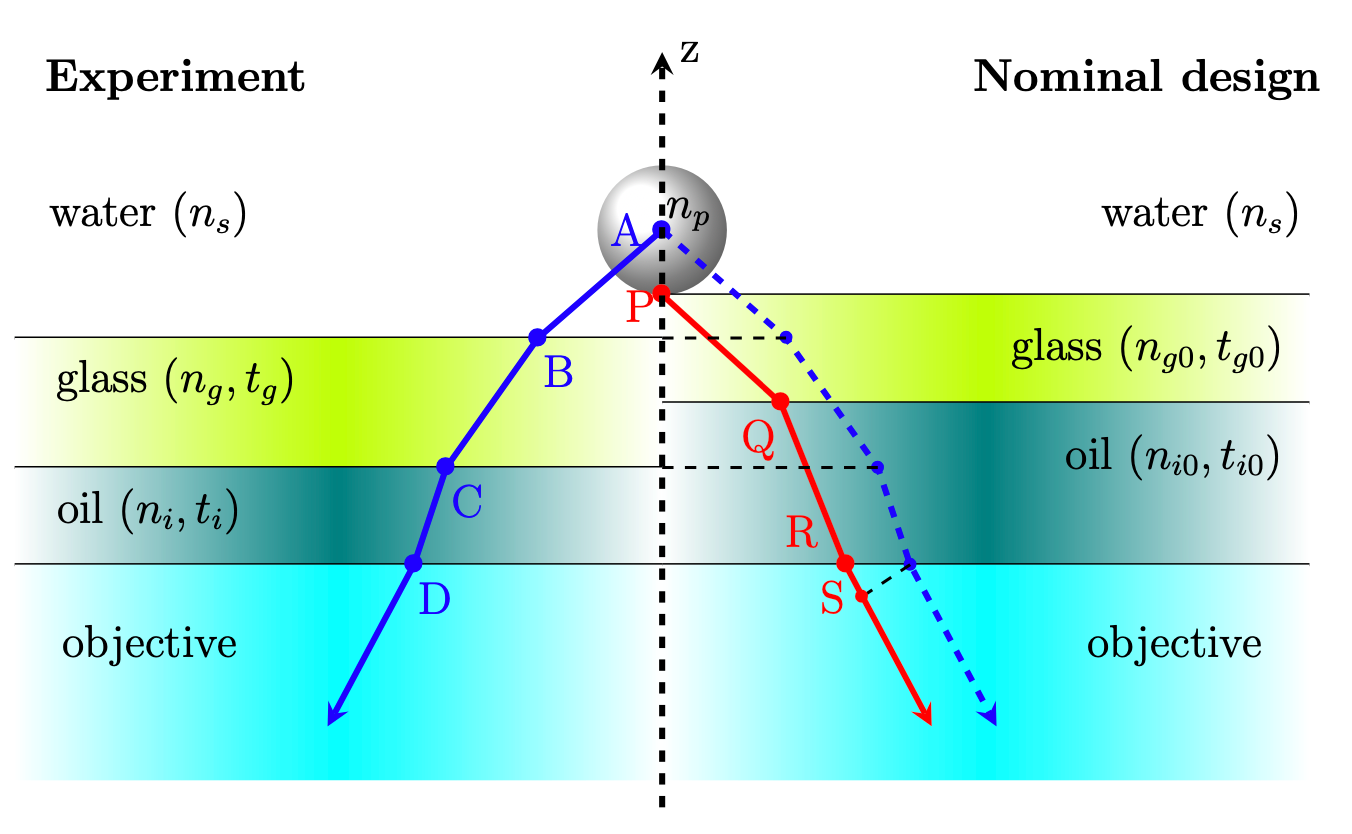}
\caption{Optical rays description of the aberration introduced for a microscope objective working in actual (left) or design (right) conditions.}  
\label{fig:PhaseAberr}
\end{figure}
We can write the thickness of the oil immersion layer as a function of the other system parameters as \cite{Haeberle:03}:

\begin{eqnarray}
t_i=z_p-z_f +n_i\left(\frac{t_{g0}}{n_{g0}}-\frac{t_{g}}{n_{g}}+\frac{t_{i0}}{n_{i0}}-\frac{z_p}{n_s}\right).
\end{eqnarray}
The optical path difference for the scattered dipolar field  $\Lambda_s=\Lambda(\theta)=(PQRS)-(ABCD)$ between the experiment and nominal design follows as (see also Fig. \ref{fig:PhaseAberr}):
\begin{eqnarray}
\nonumber
\Lambda(\theta)=&z_p\sqrt{n_{s}^2-n_i^2\sin^2\theta}+n_i \cos \theta t_i-t_{i0} \sqrt{n_{i0}^2-n_i^2\sin^2\theta}\\
&+ t_g \sqrt{n_g^2-n_i^2\sin^2\theta} -t_{g0} \sqrt{n_{g0}^2-n_i^2\sin^2\theta} 
\end{eqnarray}
Optical path aberrations for the reflected and transmitted fields at angle $\theta_m$ are accounted by including the following expressions:
\begin{eqnarray*}
\Lambda_r(\theta_m)&=&2\left(n_i t_i\cos \theta_m -t_{i0} \sqrt{n_{i0}^2-n_i^2\sin^2\theta_m}
+ t_g \sqrt{n_g^2-n_i^2\sin^2\theta_m} -t_{g0} \sqrt{n_{g0}^2-n_i^2\sin^2\theta_m} \right)\\
\Lambda_t(\theta_m)&=&z_p\sqrt{n_{s}^2-n_i^2\sin^2\theta_m} + n_i t_i\cos \theta_m -t_{i0} \sqrt{n_{i0}^2-n_i^2\sin^2\theta_m} \\
&&+ t_g \sqrt{n_g^2-n_i^2\sin^2\theta_m}-t_{g0} \sqrt{n_{g0}^2-n_i^2\sin^2\theta_m}
\end{eqnarray*}
Given that the detected intensity depends on the phase difference between the scattered and reference fields; the scattered field through the objective is updated including a phase shift $\Lambda_{tot}=\Lambda_s(\theta)+\Lambda_t(\theta_m)-\Lambda_r(\theta_m)$.  That is 
\begin{eqnarray*}
&&I_0\rightarrow \frac{in_i^{3/2}k_0^3}{n_s M^2} \frac{A_2}{2A_1}e^{ik_0\Lambda_m(\theta_m)}\int_0^{\theta_{det}} \left(\tau_p +\frac{k_s}{w_s}\tau_s \right) e^{ik_0\Lambda(\theta)}  (\cos\theta)^{\frac{3}{2}} \sin \theta  \mathrm d\theta \\
&&\Lambda_m(\theta_m)=\Lambda_t(\theta_m)-\Lambda_r(\theta_m)\\
&&\quad \quad=z_p\sqrt{n_{s}^2-n_i^2\sin^2\theta_m}\\
&&\quad \quad \quad-(n_it_i \cos\theta_m -t_{i0} \sqrt{n_{i0}^2-n_i^2\sin^2\theta_m} 
+ t_g \sqrt{n_g^2-n_i^2\sin^2\theta_m}-t_{g0} \sqrt{n_{g0}^2-n_i^2\sin^2\theta_m})
\end{eqnarray*} 

\paragraph{Numerically computed contrast signal}
Finally, we reconstruct the contrast signal from the detected intensity as the incoherent sum of the reflected plus scattered fields resulting from the K{\"o}hler excitation within the illumination numerical aperture $\mathrm{NA_i}=n_i\sin\theta_m^{max}$  
\begin{eqnarray}
I_{det}=\sum_{\theta_m\le \theta_m^{max}} \vert \vec E_r(\theta_m)+\gamma\vec E_{s}\vert^2 \label{eq:contrast}
\end{eqnarray}
We add an additional factor $\gamma$ as function of NA$_{i}$ to take into account detection efficiency of the imaging system together with discrepancies between the model and experiment in the form of differences in the beam illumination profile at the BFP of the objective, which impact the effective reflectivity (\S \ref{sec:SIreflectance}), differences in the effective scattering cross-section due to the increase in resolution (\S \ref{sec:resolution}), and deviations in the scattering contributions from the dipole approximation depending on the particle size and material (\S \ref{sect:DipoleOrient}).

\begin{figure}[H]
    \centering
    \includegraphics{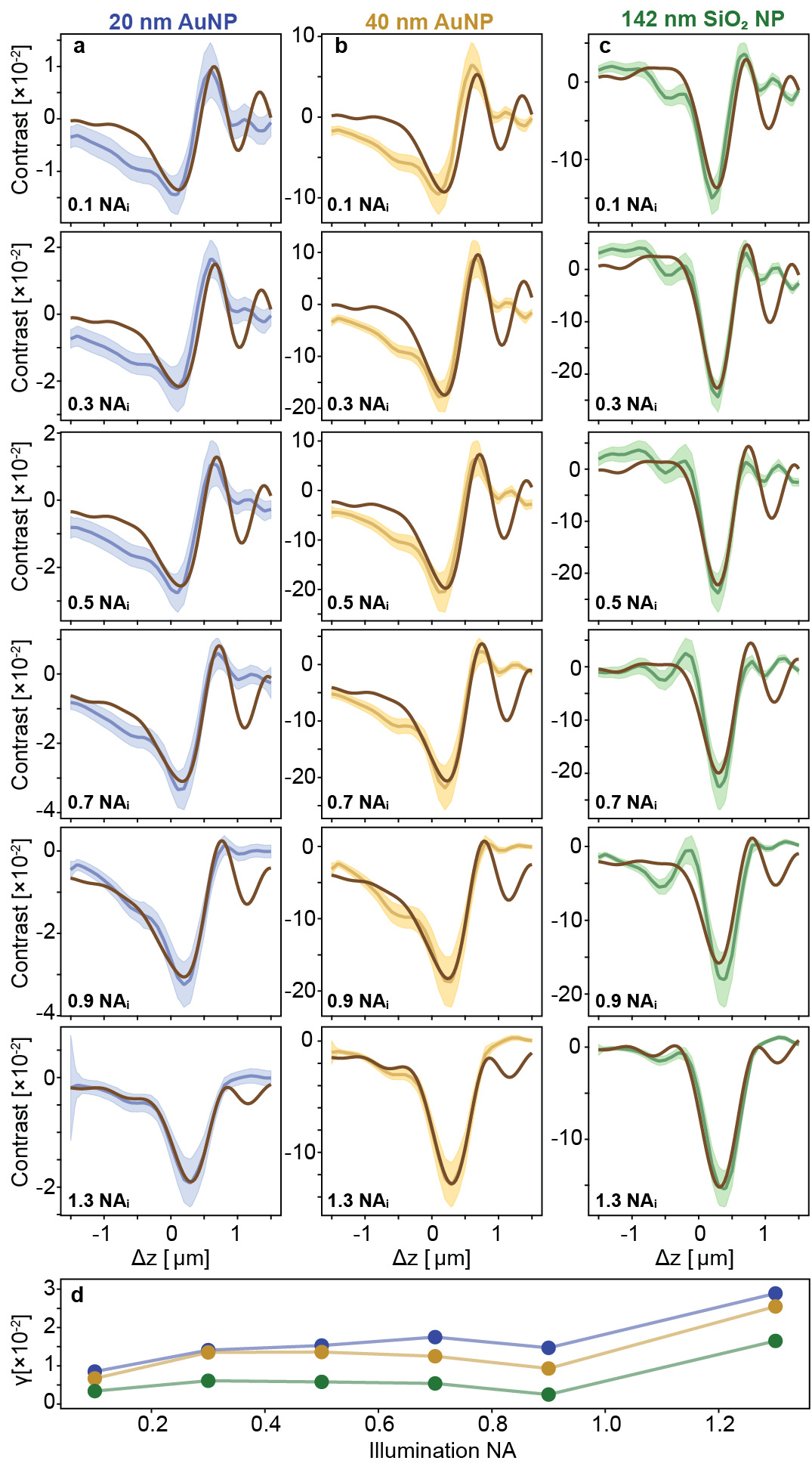}
    \caption{Partially coherent imaging model for individual NPs as a function of defocus. (a-c) Contrast defocus curves as a function of NA$_{i}$ for 20 nm AuNP, 40 nm AuNP and 142 nm SiO$_2$ NPs, respectively. Solid curves with shaded area correspond to the experimental NP ensemble average $\pm$ one standard deviation. Overlaid brown curves correspond to fits to the partially coherent imaging model. (d) Scattering field amplitude correction for the different NPs and illumination NAs. Nominal parameters are $n_{g0}=1.5$, $t_{g0}=\SI{170}{\micro \meter}$, $n_{i0}=1.5$, $t_{i0}=\SI{100}{\micro\meter}$. Fitting parameters common to all measurement are $n_{g}=1.502$, $t_{g}=\SI{174}{\micro\meter}$, $n_{i}=1.5007$. An axial offset $z_o=-71$ nm is also applied to all fitting curves.}
    \label{fig:SImodel}
\end{figure}

\paragraph{Results}
 Fig. \ref{fig:SImodel} (a-c) show the experimental results of the detected 20nm AuNP, 40nm AuNP and 142 nm SiO$_2$ contrast as a function of the axial defocus for different coherence parameters. The solid curves with shaded area correspond to the ensemble average $\pm$ one standard deviation, as shown in the main text Fig. \ref{fig:zscan1}. The overlaid brown curves correspond to a global fit to the partially coherent imaging model for all three particles and over the different degrees of partial coherence at once. With the exception of some high-frequency oscillations away from the focus, the experimental results show excellent agreement with the imaging model within $\pm$ one standard deviation of the ensemble average. We attribute this discrepancy to the spatial pixel averaging used to determine the average experimental contrast in the defocus curves. Our experimental results together with the simulations demonstrate that increasing the degree of partial coherence enhances the contrast of scattering particles. This trend is partially captured by the fit parameter $\gamma$ (Eq. \ref{eq:contrast}), whose function is to compensate for the limitations in the model with respect to the experiment (Fig. \ref{fig:SImodel}d).

\section{Origin of the contrast enhancement in a partially coherent imaging system}
To understand the complex dependence of the signal contrast on the degree of partial coherence, we decomposed the main contributing parameters for an interferometric microscope configured in a reflection geometry. For sake of simplicity, we isolated how partial coherence affects each parameter individually, whilst keeping the rest fixed. In terms of parameters, we only considered the reflectance, the imaging resolution, and the scattering emission profile under the dipole approximation. 

Given that we only want to outline the general trends rather than perform a comprehensive study, we have excluded any effects due to differences in phase transfer function, differences in phase between polarisation states, optical aberrations, and additional reflections from other interfaces in the imaging system.

\subsection{Effect of reflectance on the scattering contrast scaling as a function of degree of partial coherence} \label{sec:SIreflectance}
To determine whether the difference in signal contrast results from the change in reflectance, $R$, as the degree of partial coherence increases, we analysed its effect in the absence of any other contributions. To derive the dependence of contrast on $R$, $C(R)$, we first express the measured intensity from an interferometric-based detection as: 
\begin{equation}
    I_{det}\propto|E_R + E_s|^2 = |E_i|^2 (R +\sigma + 2\sqrt{R}\sqrt{\sigma}\cos{\theta} ), \label{Idet}
\end{equation}
with $E_R$ and $E_s$ being the reflected and scattered field, $\sigma$ the scattering cross-section of the nanoparticles, and $\theta$ the phase difference. Assuming that for all NPs considered in this work, $\sigma$ is much smaller than $R$ for the glass-water interface, and normalising the detected intensity by $E_R^2$,  we can now express the measured contrast as: 
 \begin{equation}
     C(R)= \frac{2}{\sqrt{R}}\sqrt{\sigma}\cos{\phi}. \label{C_R}
 \end{equation}
 
\begin{figure}[H]
    \centering
    \includegraphics{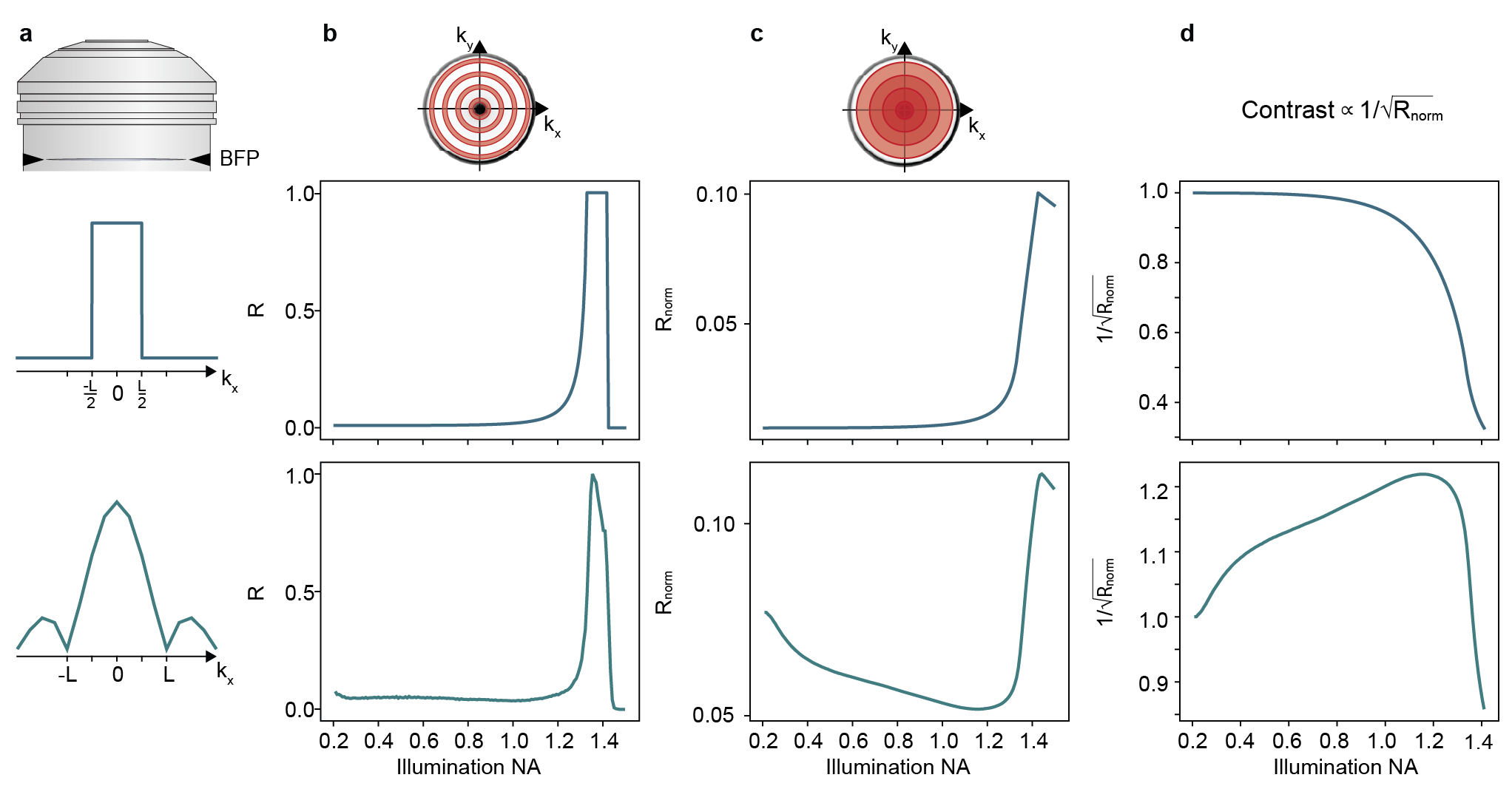}
    \caption{Effect of partial coherent illumination on the effective reflectance from the glass water interface. (a) Schematic illustrating two different implementations of partial coherent illuminations by shaping the beam profile at the BFP. Middle row: simulated data resulting from a top hat illumination; bottom column: experimentally retrieved data from the sinc-like illumination profile at the BFP. (b) Reflectance from the water/glass interface within each spatial frequency, retrieved from integrating the power at each ring from the BFP image, expressed as a function of NA$_{i}$.  (c) Effective reflectance from the water/glass interface over the total illuminated spatial frequency space expressed as a function of NA$_{i}$. (d) Effective contrast scaling caused by changes in the effective reflectance as a function of NA$_{i}$.}
    \label{fig:reflectance}
\end{figure}

The above expressions shows that the amount of light reflected at the coverslip-medium interface directly influences the contrast. Given that $R$ is a function of the incoming angle, the signal contrast depends on the NA$_{i}$ and thus the coherence parameter. Fig. \ref{fig:reflectance} shows how the reflectance changes for increasing coherence parameter. The first row shows the simulated case for flat top illumination in the backfocal plane, as shown in Fig. \ref{fig:reflectance}(a). Column (b) shows the reflectance when sampling the NA$_{i}$ as slices corresponding to single k-vectors of increasing size. In this case, the reflectance rapidly rises at the NA$_{i}$ corresponding to the critical angle and drops at the NA limits of the objective, in agreement with Fresnel coefficients derived for the glass-water interface. Column (c) shows the more experimentally relevant case of sampling the NA$_{i}$ as circular areas of increasing radii. In this case, $R_{norm}$ describes the average reflectance upon including all k-vectors contributions inside the given BFP area. As in the prior case, the reflectance increases when the NA$_{i}$ approaches the critical angle, although to a smaller extent. Assuming all other variables remain constant in Eq. \ref{C_R}, the experimental contrast can be estimated by taking the inverse square root of $R_{norm}$, as shown in (d). The decay in contrast for increasing NA$_{i}$ coincides with the experimental results of prior work \cite{Mazaheri2024ISCATCoherence, Avci2017PupilMicroscopy}.

In contrast to prior work, our platform relay images the flat top illumination from the MMF to the object plane, which results in a sinc-shaped illumination profile at the BFP. Consequently, the reflectance measured experimentally from the relay-imaged BFP is composed of the superposition of the beam profile and the reflectance from the glass-water interface derived in the first row. This is evident in (b) and (c), where $R$ rapidly rises at the critical angle but slightly decreases at larger NA$_{i}$ values. As a result of the sinc-shaded profile, the higher angular illumination components, which inherently have higher reflectance values, are weighed less compared to the flat-top illumination. This suppresses the decrease in contrast due to higher effective reflectance, whilst simultaneously decreasing coherent artifacts.  Moreover,  our experimental configuration leads to an overall contrast enhancement as the NA$_{i}$ increases, with a maximum occuring at 1.2, a reversal of the trend observed when illuminating a flat-top illumination profile at the BFP. 

\subsection{Effect of resolution increase on the scattering contrast scaling as a function of degree of partial coherence} \label{sec:resolution}
To determine whether the difference in signal contrast results from the increase in resolution as the degree of partial coherence increases, it suffices to express the partial coherent diffraction limit as $(1+s)^{-1}\rho$, where $\rho=\lambda/NA$, which is derived under the assumption of weak object optical transfer function -- meaning the absorption and phase of the NPs are sufficiently small. This leads to the well-known expressions of the diffraction limit for the coherent ($\rho$) and incoherent ($\rho/2$) cases. Experimentally we can assume that the full-width-at-half-maximum of the diffraction limit as a measure of $(1+s)^{-1}\rho$, so  the diffraction limited area as a function of degree of partial coherence can be expressed as $A(s) =  \pi (1+s)^{-2}\rho^{2}/4$. If so, we can describe the amount of scattered photons in this area per unit time as:

\begin{equation}
    P_{s}(s)=\gamma \frac{\sigma}{A(s)} P_{i}
    \label{Ps_rel}
\end{equation}

where $\gamma$, and $P_{i}$ correspond to the collection efficiency of the microscope, and the incident number of photons per area per unit time \cite{Zuo2020TransportTutorial}. If we then normalise the total amount of photons with respect to the coherent case we arrived at the following expression:

\begin{equation}
    \frac{P_{s}(s)}{P_{s}(0)}=(1+s)^2.
    \label{Ps_s}
\end{equation}

However, because the contrast signal results from an interferometric-based detection, we can approximate the number of photons in this term as $P_{int}=2\sqrt{P_{s}P_{r}}$. Assuming that the effective reflectivity of the glass-water interface remains the same, and the degree of partial coherence does not influence any other parameters, the expected contrast signal enhancement from an increase in resolution depends linearly on $(1+s)$, i.e. the degree of partial coherence. These back-of-the-envelope calculations lead to an up to a two-fold increase in contrast caused by the increase in resolution.  

\subsection{Dipole orientation and scattering efficiency as a function of degree of partial coherence}
\label{sect:DipoleOrient}
\begin{figure}[H]
    \centering
    \includegraphics{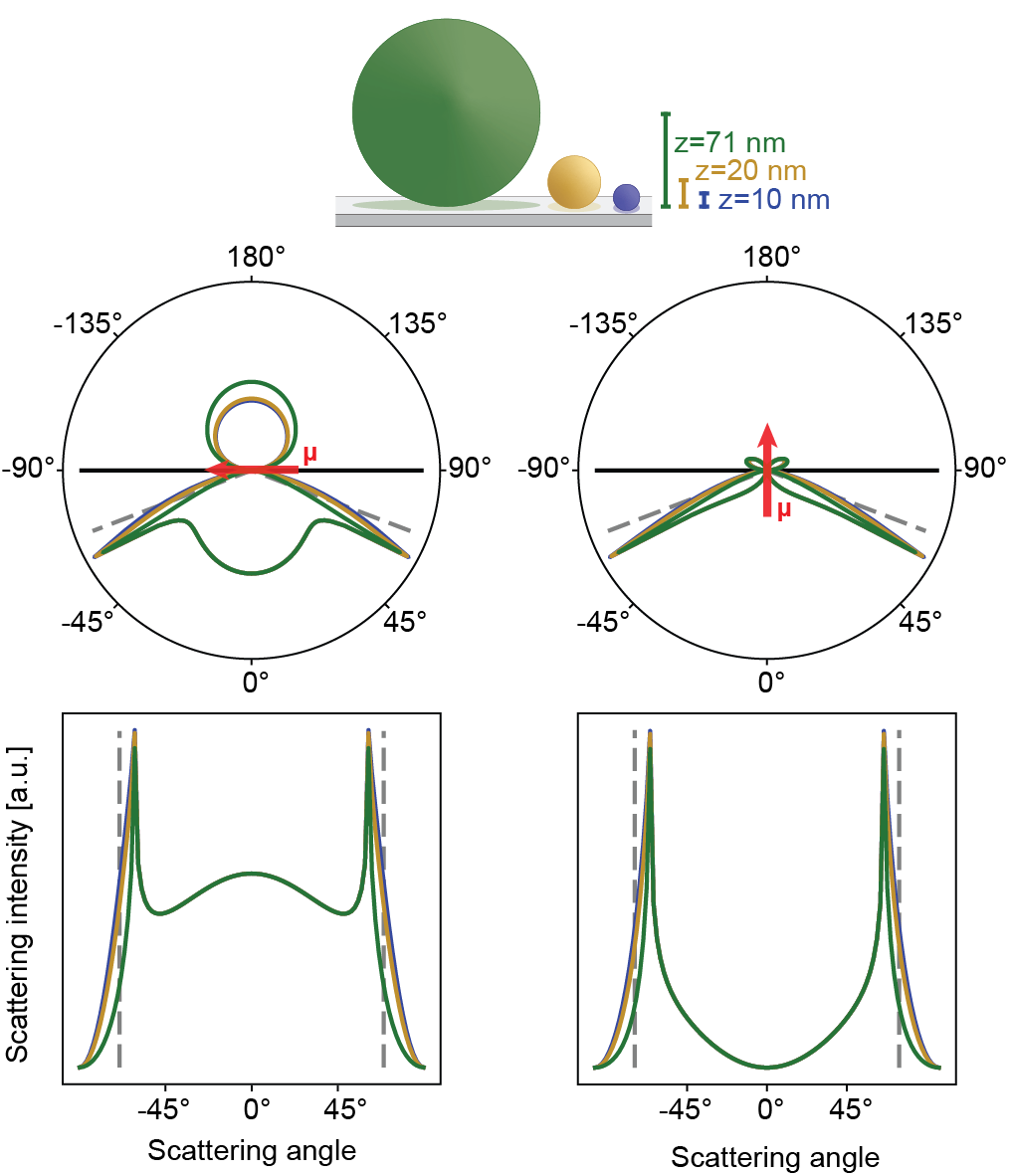}
    \caption{Effect of partial coherent illumination on scattering emission profile. Scattering emission profile for a 2D dipole located on the glass-water interface 20 nm AuNPs, 40 nm AuNPs, 143 nm SiO$_{2}$ NPs illuminated at low (left) and high (right) degrees of partial coherence. Dashed vertical lines represent the angle of collection denoted by the 1.42 NA detection objective. }
    \label{fig:scatt_profile}
\end{figure}

To determine whether the difference in signal contrast is affected by the collection efficiency caused by the change in dipole excitation from low to high NA of illumination,\cite{Avci2017PupilMicroscopy} we simulated the scattering emission profile of different particles at the glass-water interface under the dipole approximation (Figure \ref{fig:scatt_profile}). Specifically, the different particle species were approximated as point sources positioned at a distance equal to the particle's radius away from the interface. For small NA$_{i}$s, corresponding to a low degree of partial coherence, we assumed that the incident light preferentially excites dipoles parallel to the interface in the NPs, represented by a horizontal dipole. In contrast, for larger NA$_{i}$s, the incident light composed of higher spatial frequencies leads to excitation of partially vertical dipoles in the NPs, represented in the plots as a vertical dipole.

The upper row of plots in Fig. \ref{fig:scatt_profile} shows the backwards (-90° to 90°) and forwards (-180° to -90° and 90° to 180°) scattering of the three particles, under either a perfectly horizontal (left) or vertical (right) dipole orientation, respectively. The lower row shows the corresponding backwards scattering contribution, with the gray dashed lines delineating the maximum collection angle from the detection objective (1.42 NA, oil immersion). Despite that for the vertical dipole case, a larger portion of light is backscattered compared to the horizontal dipole case; the total scattered intensity is lower for vertical dipoles, resulting in a decrease in collected light of 11\%, 13\%, and 19\% for 20 nm AuNPs, 40 nm AuNPs, and 142 nm $\mathrm{SiO}_2$ NPs, respectively, compared to the horizontal dipole case. This trend of decreasing scattering intensity, and thus, lower contrast as a function of degree of partial coherence, was not observed experimentally. As such, we can rule out this contribution as being the predominant factor affecting the contrast modulation. Furthermore, even at the highest NA$_{i}$ examined, a combination of horizontal and vertical dipoles is present, rather than purely vertical dipole contribution. Lastly, describing the emission as a pure dipole point source is not completely accurate, particularly for the 142 nm silica particles, where higher-order modes become increasingly relevant.

\section{Supplementary figures for EV-based measurements}
\begin{figure}[H]
    \centering
    \includegraphics{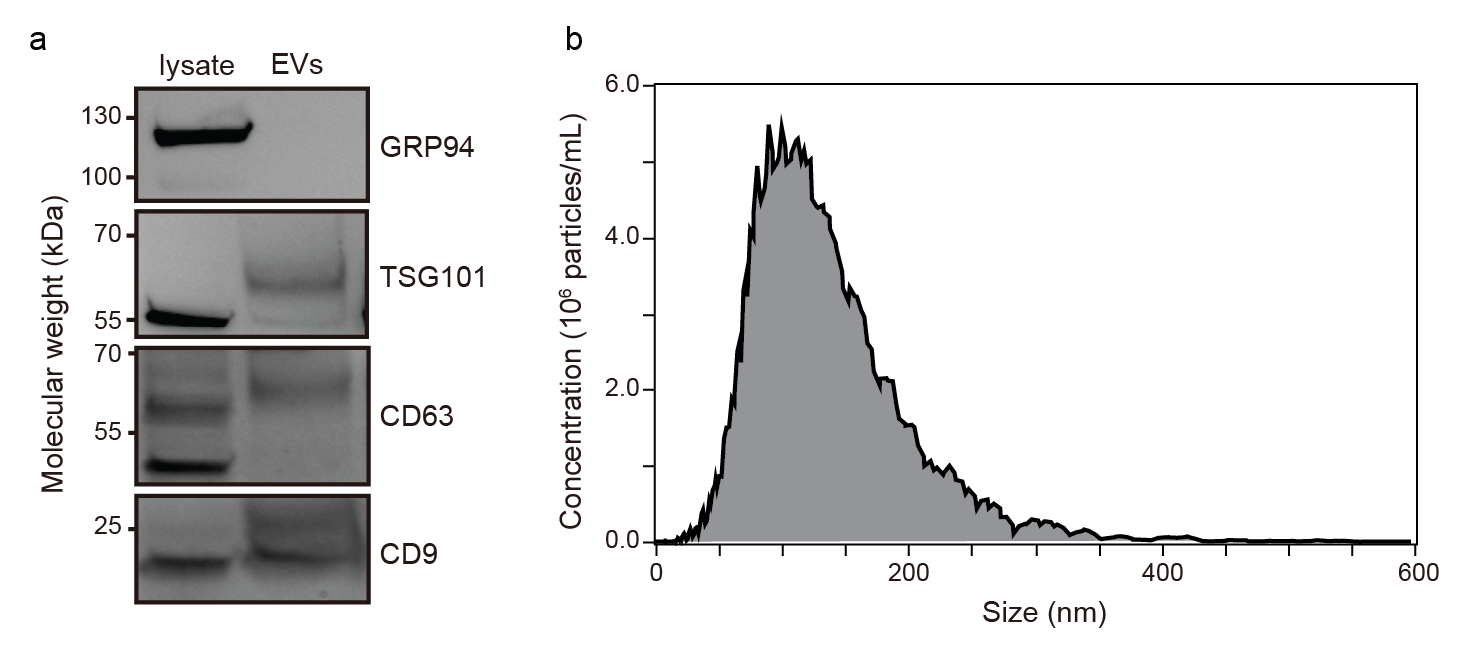}
    \caption{H358 EV characterisation. (a) Western blot analysis against specific (TSG101, CD63, CD9) and non-specific (GRP94) EV markers. (b) Size and concentration determination via nanoparticle tracking analysis. }
    \label{fig:SI_EVcharacterisation}
\end{figure}

\begin{figure}[H]
    \centering
    \includegraphics[width=0.5\linewidth]{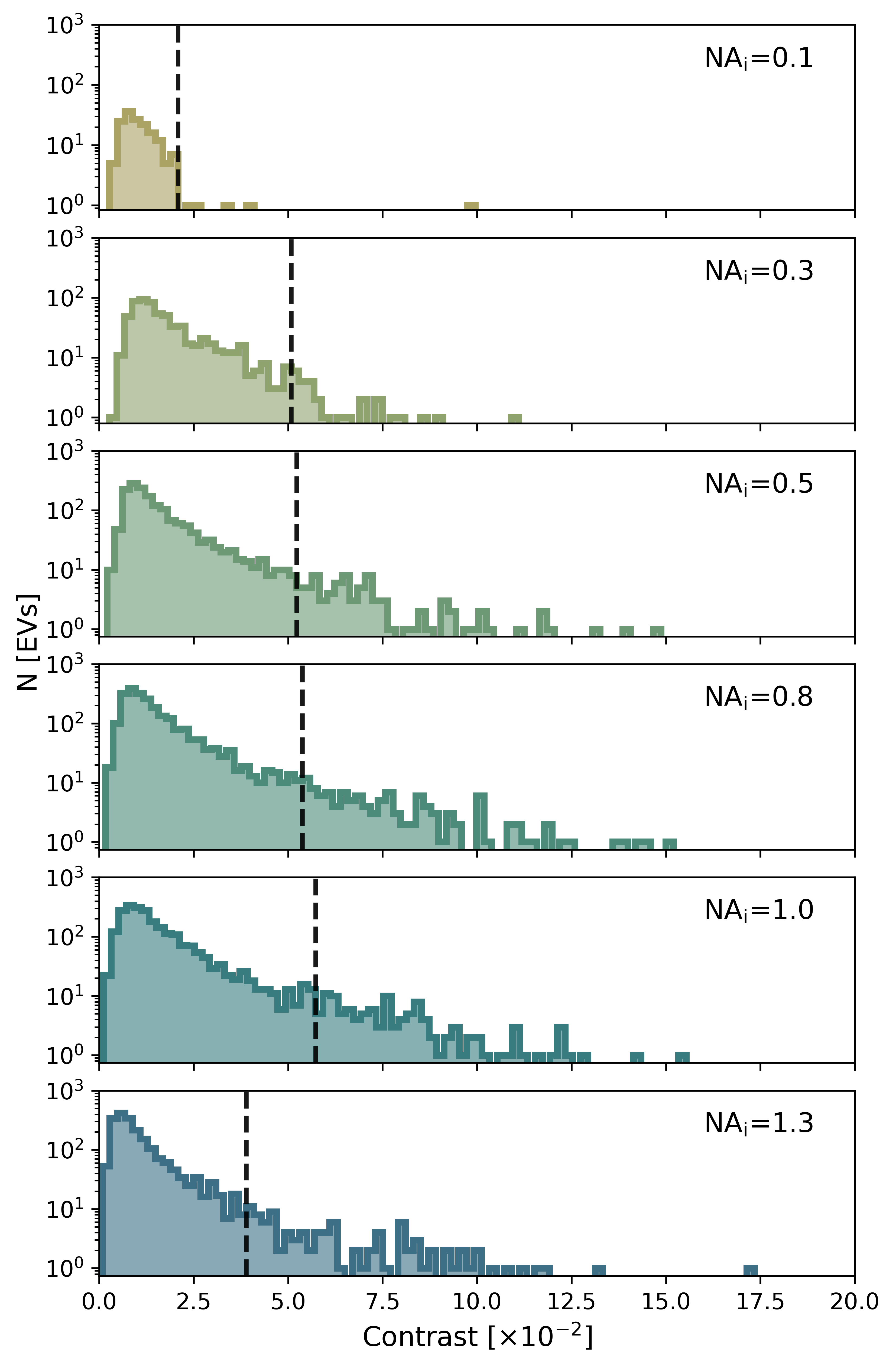}
    \caption{Distribution of the maximal contrast of all detected EVs as a function of the degree of partial coherence. Dashed lines indicate the 95th percentile value. Increasing from the smallest to the largest NA$_{i}$, the counts of considered EVs are 160, 683, 1746, 2501, 2491, 2098.}
    \label{fig:SI_EVcontrastDist}
\end{figure}

\begin{figure}[H]
    \centering
    \includegraphics{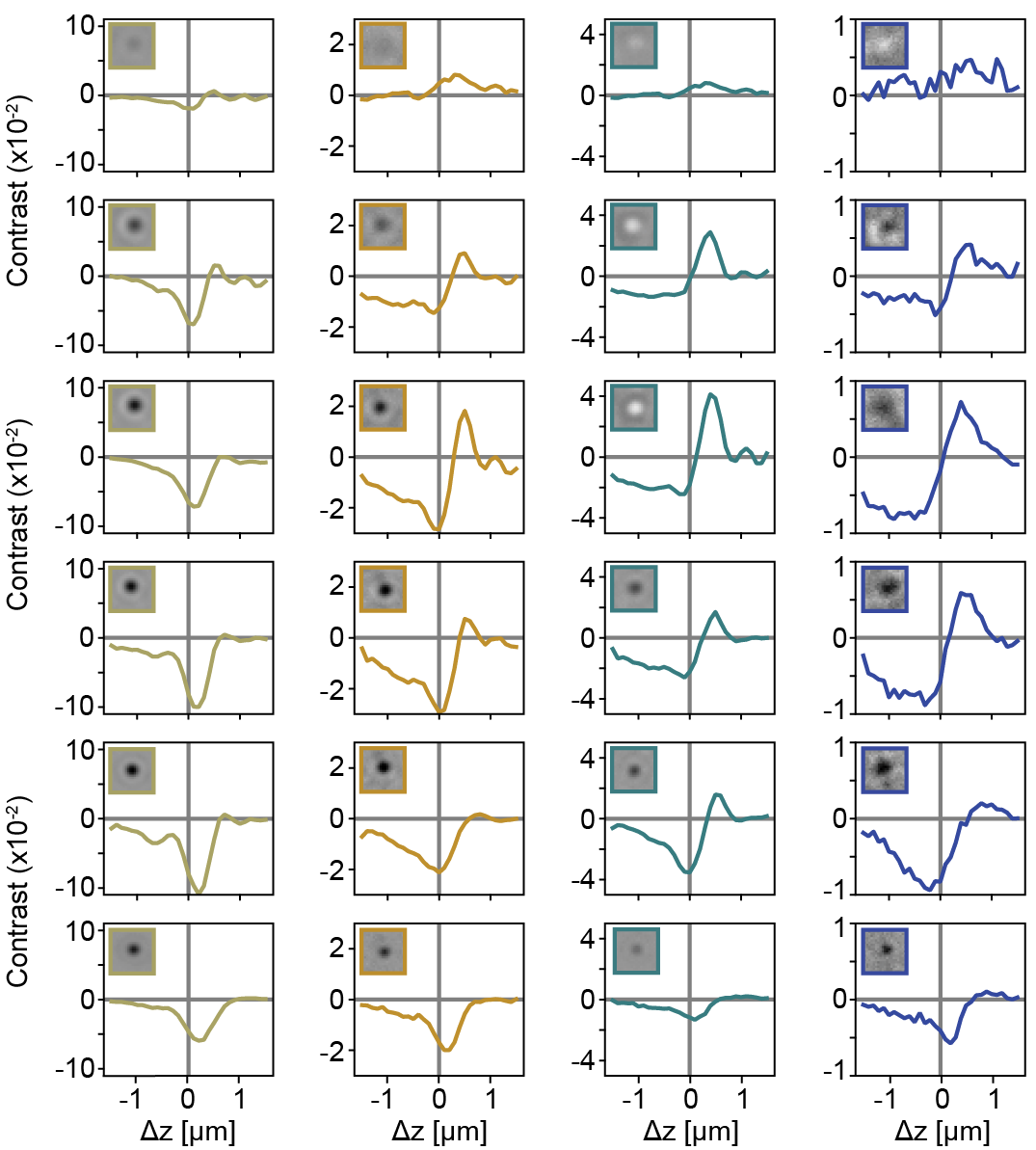}
    \caption{Contrast defocus curves as a function of the degree of partial coherence for the four representative EV particles shown in Figure 5b-d. Inset: zoom-in image of the PSF of the EVs at the focus position that maximises the absolute signal contrast value.}
    \label{fig:SI_allsingleEVs}
\end{figure}

\section{Supplementary figure for single protein sensing and illumination beam engineering}

\begin{figure}[H]
    \centering
    \includegraphics{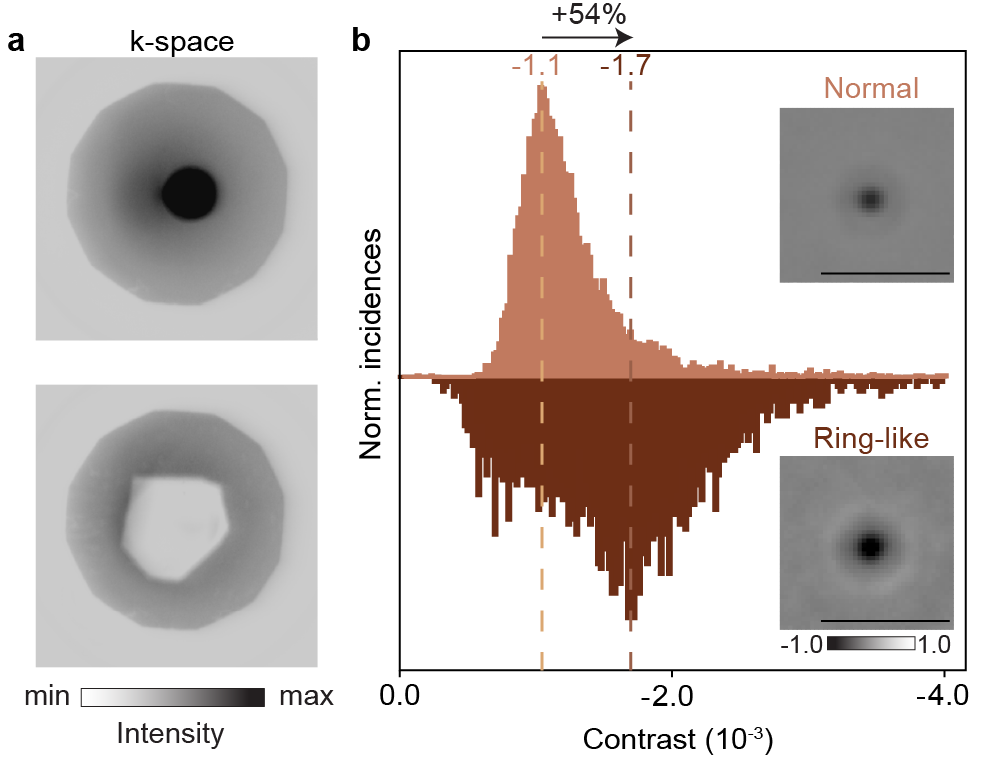}
    \caption{Illumination beam engineering enhances the signal contrast. (a) BFP image upon a partial coherent illumination corresponding to NA$_{i}$=0.9 without (top) and with (bottom) illumination beam engineering. For beam profile engineering, an additional aperture stop was placed in the illumination module of the setup to create a ring-like partially coherent illumination profile. (b) Corresponding contrast distribution of the detected TG binding events comparing both illumination profiles, with vertical dotted lines indicating the maxima of each distribution. The ring-like illumination leads to a 54\% increase in contrast signal from which a second weaker population is discernible, corresponding to the TG monomer. Inset: ensemble-averaged particle PSF for all binding events. Scale bar: $\SI{1}{\micro\meter}$}
    \label{fig:SI_BeamEng}
\end{figure}
\end{document}